\documentclass[universe,article,accept,pdftex,moreauthors]{Definitions/mdpi} 

\usepackage{longtable}
\usepackage{rotating}
\usepackage{pdflscape}
\usepackage{soul}

\firstpage{1} 
\pubvolume{1}
\issuenum{1}
\articlenumber{0}
\pubyear{2024}
\copyrightyear{2024}
\datereceived{ } 
\daterevised{ } % Comment out if no revised date
\dateaccepted{ } 
\datepublished{ } 
\hreflink{https://doi.org/} % If needed use \linebreak
\Title{High-redshift quasars at $z \geq 3$ - III. Parsec-scale jet properties from VLBI observations}

\TitleCitation{High-redshift quasars at $z \geq 3$ - III. Parsec-scale jet properties from VLBI observations}

\Author{Shaoguang Guo $^{1,2,3}$\orcidA{}, Tao An $^{1,2,3,4}$*\orcidB{} , Yuanqi Liu $^{1}$\orcidC{}  , Chuanzeng Liu $^{1,2,5}$\orcidD{}, Zhijun Xu$^{1}$\orcidE{}, Yulia Sotnikova $^{6,7,8}$\orcidF{},  Timur Mufakharov $^{6,7,8}$\orcidG{} and Ailing Wang$^{9}$ $\orcidH{}$}

\AuthorNames{Shaoguang Guo, Tao An, Yuanqi Liu et al.}

\AuthorCitation{Shaoguang Guo, Tao An, Yuanqi Liu et al.}
\address{%
$^{1}$ \quad Shanghai Astronomical Observatory,  Chinese Academy of Sciences, 80 Nandan Road, Shanghai 200030, China;\\
$^{2}$ \quad University of Chinese Academy of Sciences, 19A Yuquanlu, Beijing 100049, China; \\
$^{3}$ \quad Key Laboratory of Radio Astronomy and Technology, Chinese Academy of Sciences, A20 Datun Road, Chaoyang District, Beijing 100101, China;\\
$^{4}$ \quad Xinjiang Astronomical Observatory, Chinese Academy of Sciences, 150 Science 1-Street, Urumqi, Xinjiang 830011, China ;\\
$^{5}$ \quad School of Physical Science and Technology, ShanghaiTech University, Shanghai 201210, China;\\
$^{6}$ \quad Special Astrophysical Observatory of RAS, Nizhny Arkhyz, 369167, Russia; \\
$^{7}$ \quad Kazan Federal University, 18 Kremlyovskaya St, Kazan 420008, Russia;\\
$^{8}$ \quad Institute for Nuclear Research, Russian Academy of Sciences, Kutuzova Embankment 10, St. Petersburg 191187, Russia;\\
$^{9}$ \quad Key Laboratory of Particle Astrophysics, Institute of High Energy Physics, Chinese Academy of Sciences, Beijing 100049, China;
}

\corres{Correspondence: antao@shao.ac.cn}

\abstract{High redshift active galactic nuclei (AGN) provide key insights into early supermassive black hole growth and cosmic evolution. This study investigates the parsec-scale properties of 86 radio-loud quasars at z $\geq$ 3 using very long baseline interferometry (VLBI) observations. Our results show predominantly compact core and core-jet morphologies, with 35\% unresolved cores, 59\% core-jet structures, and only 6\% core-double jet morphology.
Brightness temperatures are generally lower than expected for highly radiative sources. The jet proper motions are surprisingly slow compared to lower-redshift samples. We observe a high fraction of young and/or confined peak-spectrum sources, providing insights into early AGN evolution in dense environments during early cosmic epochs. The observed trends may reflect genuine evolutionary changes in AGN structure over cosmic time, or selection effects favoring more compact sources at higher redshifts. These results stress the complexity of high-redshift radio-loud AGN populations and emphasize the need for multi-wavelength, high-resolution observations to fully characterize their properties and evolution through cosmic history.}

\keyword{galaxies: active – galaxies: high-redshift – galaxies: jets – quasars: general} 

\begin{document}

\section{Introduction}

The study of high-redshift active galactic nuclei (AGN) provides unique insights into not only the early evolution of supermassive black holes (SMBHs) and their host galaxies but also fundamental physics under extreme conditions during a critical period in cosmic history \citep{2006ARA&A..44..415F, 2012Sci...337..544V, 2012RPPh...75l4901V,2020ARA&A..58...27I}. While multi-wavelength observations have revealed much about black hole accretion processes,  very long baseline interferometry (VLBI) observations of relativistic jets offer direct probes of the innermost regions around these early black holes, providing crucial measurements of their energetic output and feedback mechanisms \citep{2012ARA&A..50..455F, 2020NewAR..8801539H, 2019ARA&A..57..467B, 2013MNRAS.432.2818G}.

At lower redshifts, radio-loud AGN constitute approximately 10\% of the entire AGN population \citep{2002AJ....124.2364I, 2016ApJ...831..168K}. However, the evolution of this fraction with cosmic time and its dependence on physical parameters remains heavily debated. Recent studies have revealed complex trends, with some suggesting an increase in the radio-loud fraction with optical luminosity \citep{2007ApJ...656..680J}, while others find evidence for a roughly constant fraction across a wide range of redshifts and luminosities \citep{2002AJ....124.2364I,2016ApJ...831..168K}.  
The apparent contradictions in these findings highlight the challenges in disentangling genuine evolutionary effects from observational biases.

Theoretical work increasingly suggests that radio jets may play a particularly important role in galaxy evolution at high redshifts through mechanical feedback \citep{2020NewAR..8801539H}. The intense energy output from these jets can significantly impact the surrounding interstellar and intergalactic medium, potentially regulating both star formation and black hole growth in the early Universe. Understanding the physical properties of these jets is therefore essential for constraining models of galaxy formation and evolution.

Blazars \citep{1980ARA&A..18..321A}, a subset of AGN characterized by relativistic jets oriented close to our line of sight \citep{1995PASP..107..803U, 1978bllo.conf..328B}, are preferentially detected in flux-limited surveys due to Doppler boosting effects, particularly at high redshifts \citep{2011MNRAS.416..216V}. This selection bias must be carefully considered when interpreting population statistics. While blazars typically exhibit flat radio spectra \citep[e.g.][]{2014AJ....147..143H}, a surprising finding from recent surveys is that a significant fraction of high-redshift radio-emitting AGN have been found to display steep or peaked spectra \citep{2017MNRAS.467.2039C}. 

These spectral characteristics are more commonly associated with either young radio sources or those confined by dense environments \citep{2000MNRAS.319..445S,2009AN....330..120F}, such as compact steep-spectrum (CSS) and GHz-peaked spectrum (GPS) sources \citep{1998PASP..110..493O, 2009AN....330..120F, 2016AN....337....9O, 2021A&ARv..29....3O}. The study of GPS and CSS sources at high redshifts provides crucial insights into the early stages of jet development and its interaction with the host galaxy \citep[e.g.][]{1995A&A...302..317F, 2012ApJ...760...77A}.

The high prevalence of GPS and CSS sources at high redshifts raises fundamental questions about the physical conditions that may favor the formation of these compact radio sources in the early Universe. Multiple factors could contribute to this phenomenon, including higher ambient gas densities affecting jet propagation, more frequent galaxy mergers triggering AGN activity,  different accretion modes in early SMBHs, and selection effects in current surveys \citep{2016A&ARv..24...10T, 2018A&ARv..26....4M}. Understanding the relative abundance of these factors is crucial for constraining models of AGN evolution across cosmic time.

In the first paper of this series, Sotnikova et al. \citep{2021MNRAS.508.2798S} (Paper I) presented a comprehensive study of 102 high-redshift ($z \geq 3$) quasars with $S_{\rm 1.4GHz}>100$ mJy observed using the RATAN-600 telescope. 
They found that nearly 50\% of the sample exhibited peaked spectra, while only 24\% showed flat spectra typical of blazars, suggesting a significant population of young, evolving radio AGN at high redshifts. These sources offer a unique opportunity for investigate SMBHs during their formative stages. 
The powerful jet feedback mechanisms in young radio AGN \citep[e.g.][]{2008MNRAS.387..639H,2018A&ARv..26....4M} provide insights into the co-evolution of black holes and their host galaxies in the early Universe.

The second paper in our series \citep{2024Galax..12...25S} (Paper II) focused on the variability properties of these high-redshift quasars, revealing complex patterns of flux density variations across multiple frequencies. This variability analysis provided important constraints on the physical processes occurring in these distant AGN, complementing the spectral classification from the first paper.

In this third paper, we present the first systematic VLBI study of parsec-scale jet properties in a well-defined, flux-limited sample of 86 radio-loud quasars at $z \geq 3$.  Using multi-frequency VLBI observations spanning multiple epochs, we aim to address several key questions: what are the predominant morphological characteristics of jets in high-redshift quasars? how do the core properties (size, flux density, brightness temperature) of these distant sources relate to their overall spectral characteristics and variability? For sources with multi-epoch data, what can we infer about jet kinematics and evolution at high redshifts? 
How do the physical properties of GPS and CSS sources at high redshifts compare to their lower-redshift counterparts?

At the redshifts of our targets ($z = 3-5.3$), these observations probe physical scales of just a few parsec in projection\footnote{Throughout this paper, we adopt a flat $\Lambda$CDM cosmological model with H$_{0}=70$\,km s$^{-1}$ Mpc$^{-1}$, $\Omega_{\Lambda}=0.73$, and $\Omega_{\mathrm{m}}=0.27$. At $z = 3$, 1 mas angular size corresponds to a projected linear size of $\sim$7.7 pc, or $\sim$25 light years.}, providing unprecedented resolution of the jet collimation regions. This allows us to directly study the physics of relativistic jets during an epoch when the Universe was less than 2 billion years old.

By combining our VLBI analysis with previous spectral and variability studies, we aim to present a comprehensive picture of the properties of the radio-brightest high-redshift quasars, from their integrated spectra and variability characteristics to their parsec-scale jet structures. This multi-faceted approach allows us to probe the physical processes governing jet formation and propagation in the early Universe, providing valuable constraints on models of SMBH growth and AGN evolution during the crucial transition from cosmic dawn to cosmic noon.

\section{Sample} \label{sec:sample}

\subsection{Sources with VLBI data}

Our study builds upon a well-defined parent sample of 102 high-redshift quasars initially observed by Sotnikova et al. \citep{2021MNRAS.508.2798S}. This parent sample was selected based on three key criteria: redshift $z\geq 3$; 1.4-GHz total flux densities exceeding 100 mJy; declination range from $-35^\circ$ to $+49^\circ$. From this parent sample, we identified 86 sources with available VLBI data suitable for detailed structural analysis (Appendix \ref{app:sample}). 
While this selection ensures sufficient signal-to-noise ratios for reliable VLBI imaging, we note that the resulting sample cannot be considered strictly complete due to the VLBI data availability criterion. Therefore, the statistical results should be interpreted with appropriate caution.

% highest z is 5.28 J1026+2542
The redshift range of our targets ($z = 3 - 5.3$) encompasses a crucial epoch in cosmic history, spanning from the end of the Epoch of Reionization (which ended around $z \sim 5.3$) to the onset of the cosmic noon (around $z \sim 3$). This period marked by rapid evolution in both galaxy properties and black hole growth rates, provides a unique window into AGN physics . 
% The observed decrease in source count with increasing redshift aligns with cosmological expectations and the evolution of AGN populations during this transformative era.

Our sample selection process involves several important selection effects that must be carefully considered when interpreting the results. While the radio selection criterion ($S_\text{1.4GHz} > 100$ mJy) is well-defined, the requirement for spectroscopic redshifts introduces an additional selection bias, potentially favoring optically brighter sources. Of the original 102 sources in the parent sample, 16 lack VLBI detections in the Astrogeo VLBI database\footnote{Astrogeo VLBI archive, \url{http://astrogeo.org/}} . Analysis of their single-dish measurements suggests these sources may have more extended structure that is resolved out at VLBI scales, implying our VLBI-detected sample could be biased toward more compact sources.

The geodetic/astrometric nature of our VLBI observations introduces further selection effects. These observations typically target compact, bright sources optimal for geometric measurements --- a fundamentally different selection than imaging-focused surveys like the MOJAVE (Monitoring Of Jets in Active galactic nuclei with VLBA Experiments)  \citep{2016AJ....152...12L, 2018ApJS..234...12L, 2021ApJ...923...30L}, which often preferentially observe sources with complex, evolving jet structures. This distinction is particularly important when comparing our results with lower-redshift samples (Section  \ref{sec:pm}).

Table \ref{tab:sample} presents the fundamental properties of our sample. Redshifts (column 2) were adopted from the NASA/IPAC Extragalactic Database (NED), while precise coordinates are derived from the Astrogeo archive.  Radio spectral classifications are adopted from \citet{2021MNRAS.508.2798S}, and morphological classifications are based on our VLBI image analysis.  The sample spans a range of radio spectral types, from flat-spectrum sources typical of blazars to peaked-spectrum objects characteristic of young or evolving radio sources.
% Right Ascension (RA) and Declination (Dec) coordinates (column 3) are derived from the Astrogeo VLBI database, ensuring precise positional information crucial for VLBI analysis. The radio spectral type classifications (column 4) are taken from Ref. \citep{2021MNRAS.508.2798S}, while the morphological classifications (column 5) are based on our analysis of the VLBI images. 

Among our sample, several sources are particularly noteworthy warranting detailed study: 
(1) the highest-redshift source, 102623+254259 ($z = 5.3$), represents one of the most distant known radio-loud AGN. This source exhibits a one-sided core-jet morphology on milliarcsecond scales \citep{2013MNRAS.431.1314F}, with its jet oriented close to the line of sight. Its characteristics, including a large bulk Lorentz factor, high brightness temperature, and fast jet proper motion, make it an exceptional laboratory for studying relativistic jets in the early Universe \citep{2015MNRAS.446.2921F}.
(2) 160608+312504 ($z = 4.56$, corresponding to J1606+3124) shows a rare compact symmetric object (CSO) morphology at such a high redshift. Its GHz-peaked spectrum (GPS) with a rest-frame turnover frequency of 10.6-17.8 GHz provides important constraints on  radio source evolution or confinement models \citep{2022MNRAS.511.4572A}. Notably, this source is probably classified as a galaxy rather than a quasar, adding diversity to our high-redshift radio source population.

\subsection{VLBI data}

Our study utilizes VLBI data obtained primarily from the Astrogeo VLBI archive, a comprehensive database containing calibrated VLBI observational data. 
The Astrogeo archive consists primarily of astrometric and geodetic VLBI observations, which typically have shorter integration times per source compared to dedicated imaging observations. This affects the sensitivity and uv-coverage of the data, resulting in typical image noise levels ranging from 0.1 to 1 mJy~beam$^{-1}$ and a practical flux density sensitivity threshold of approximately 0.05-0.1 Jy at X-band and 0.1-0.2 Jy at S-band. This sensitivity limitation is reflected in the lower flux density cutoff observed in our sample distribution (Figure 1). Weaker sources were less frequently targeted due to correlator constraints and detectable signal-to-noise ratios.

For our sample, we collected all available VLBI images and calibrated visibility data. The observations span multiple frequencies, predominantly including 2.3 GHz (S band), 4.7/5.0 GHz (C band), 7.6/8.4 GHz (X band), 15 GHz (U band) and 24 GHz (K band) for selected sources.  This multi-frequency coverage enables detailed spectral properties and investigation of frequency-dependent effects on source structure. The X-band (8.4 GHz) observations form the core of our analysis, providing an optimal balance between resolution, sensitivity, and consistent coverage across the sample. 

\begin{figure}
    \centering
    \includegraphics[width=0.8\textwidth]{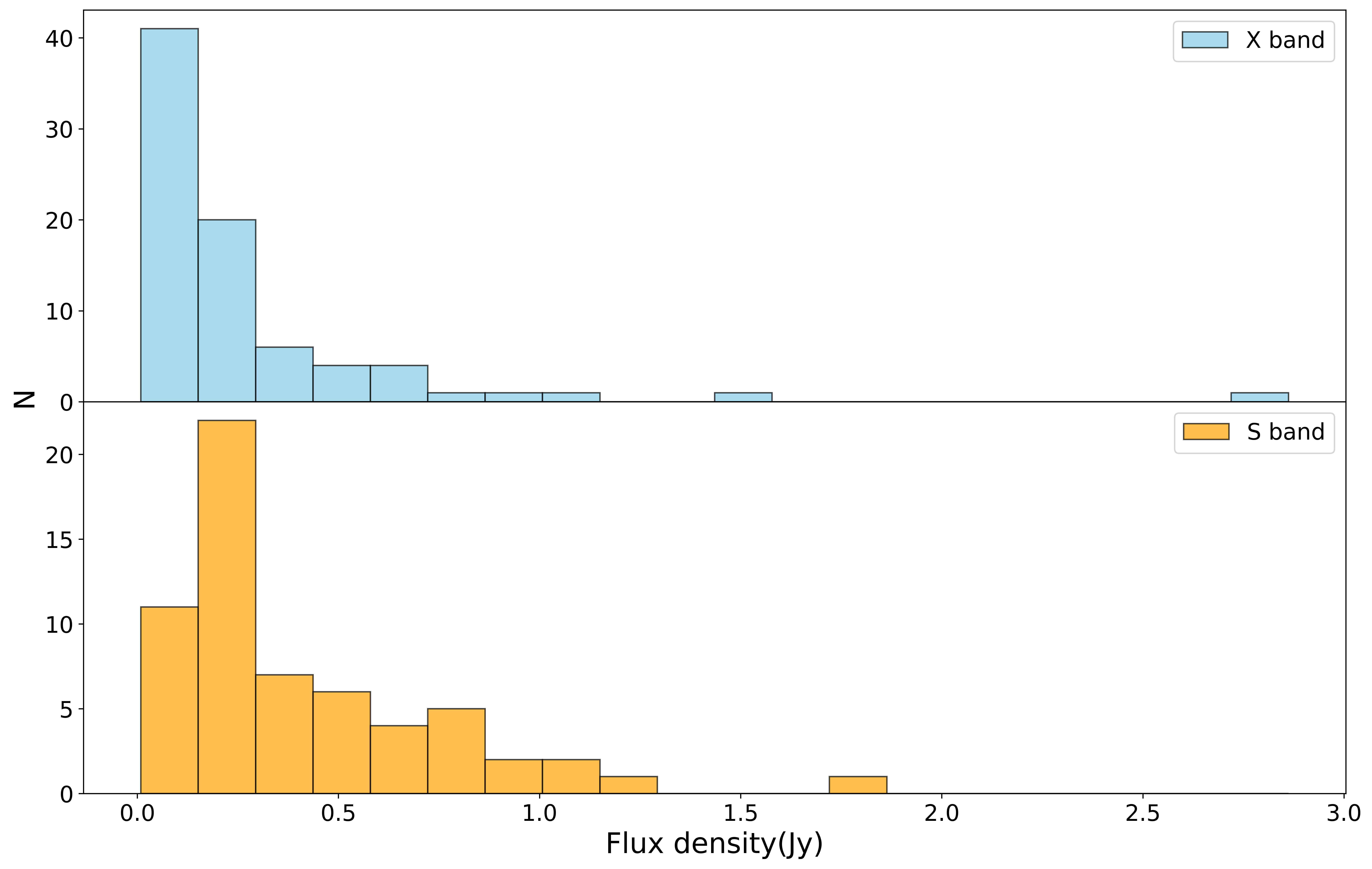}
    \caption{Histogram of the X-band and S-band VLBI flux density.}
    \label{fig:flux}
\end{figure}

Figure \ref{fig:flux} presents the VLBI flux density distribution of our sample at both X-band (upper panel) and S-band (lower panel). The X-band ($\sim$8.4 GHz) flux density distribution shows an asymmetric profile with a pronounced peak at $0.1 \sim 0.25$ Jy and a long tail extending to higher flux densities up to $\sim$3.0 Jy. The S-band ($\sim$2.3 GHz) distribution exhibits a similar asymmetric pattern but peaks at slightly higher flux densities ($0.2 \sim -0.4$ Jy).

Several key features are noteworthy. First, both bands show a sharp decline in source counts with increasing flux density, reflecting both the intrinsic rarity of bright radio sources and following the expected luminosity function of radio-loud AGN. Second, the lower flux density cutoff around 0.05-0.1 Jy in X-band and 0.1-0.2 Jy in S-band represents the sensitivity threshold of current VLBI observations rather than an intrinsic feature of the source population. Third, the different distributions between X and S bands provide insights into the spectral properties of these sources, with the systematically higher S-band fluxes indicating predominantly steep or peaked spectra.

%Figure \ref{fig:flux} shows the histogram of X-band VLBI flux density. The X-band flux density distribution shows an asymmetric profile with a pronounced peak at $0.1 \sim0.25$ Jy and a long tail extending to higher flux densities. The sharp decline in source counts with increasing flux density reflects both the intrinsic rarity of bright radio sources and follows the expected luminosity function of radio-loud AGN. A lower flux density cutoff around 0.05-0.1 Jy is evident, representing the sensitivity threshold of current VLBI observations rather than an intrinsic feature of the source population.

The observational data in our study span from the early 1990s to 2022, providing a substantial temporal baseline for many sources. Notably, 62 out of our 86 sources have $\geq 2$ epochs of observations, with some having up to 163 epochs for J0646+4451 (OH~471) \citep{2024A&A...685L..11G}, 137 epochs for J2129$-$1538. Our VLBI dataset complements and extends the RATAN-600 observations. While the RATAN-600 data covered a more recent and concentrated period from 2006 to 2019 \citep{2021MNRAS.508.2798S} and from 2006 to 2022 \citep{2024Galax..12...25S}, our VLBI dataset provides a longer historical perspective, reaching back to the early 1990s. This extensive temporal coverage is particularly valuable given cosmic time dilation effects. For our highest redshift sources, a 30-year observing baseline in the observer's frame corresponds to only about 5-7 years in the source rest frame, necessitating long monitoring periods to study structural evolution.

The observations were conducted using various VLBI arrays, predominantly the Very Long Baseline Array (VLBA), but also including the European VLBI Network (EVN) and global VLBI arrays. This diversity of arrays provides complementary coverage of different angular scales and sensitivities, enabling robust characterization of source structures across a wide range of spatial frequencies.

This VLBI dataset, combined with our previous single-dish RATAN-600 monitoring and spectral analysis, provides comprehensive view of high-redshift radio-loud AGN properties and evolution. The combination of high angular resolution, multi-frequency coverage, and extensive temporal sampling makes this dataset uniquely suited for investigating the physics of relativistic jets in the early Universe.

% flux density ratio distribution 
% TODO 
% xlabe -> $C$
\begin{figure}
    \centering
    \includegraphics[width=0.8\textwidth]{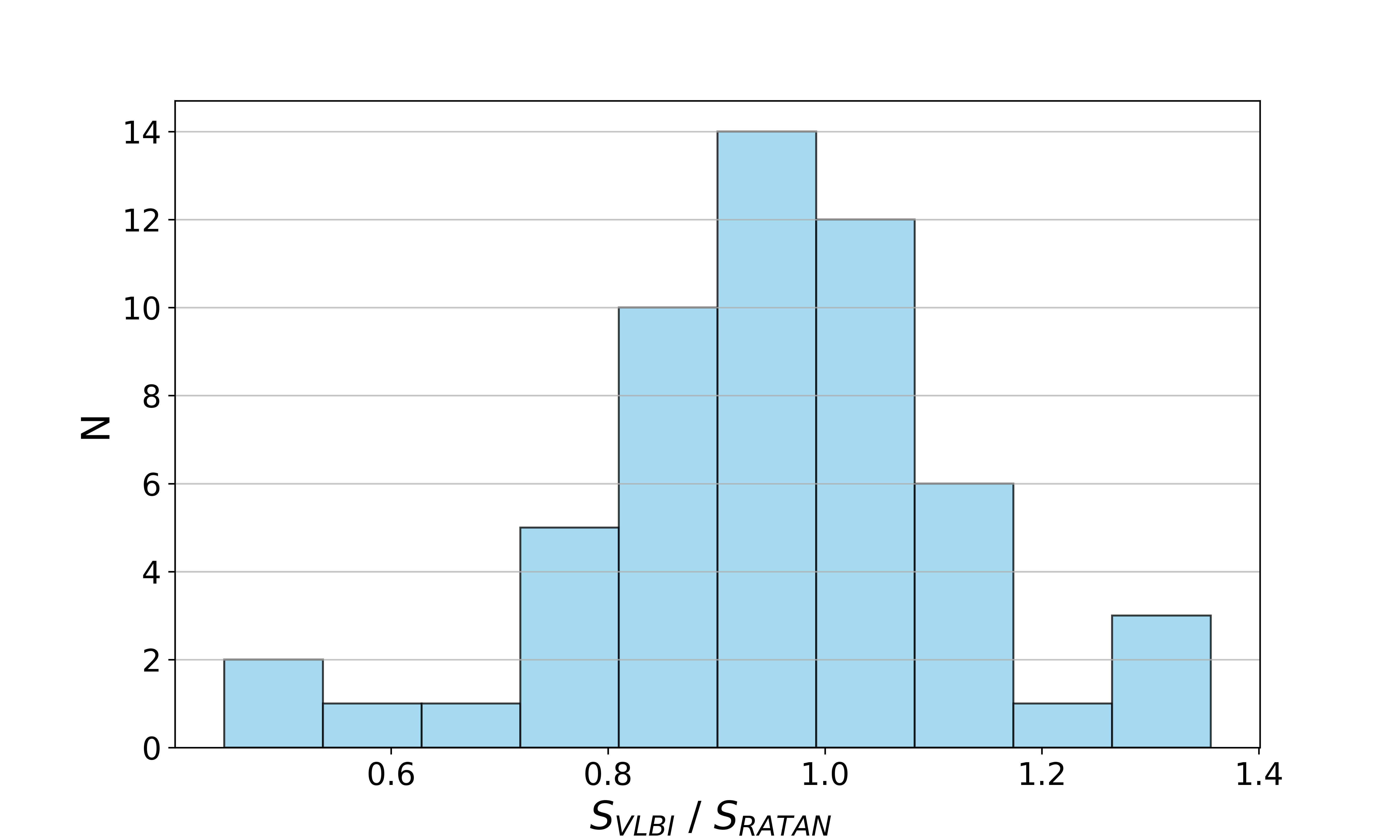}
    \caption{Histogram of the source compactness index $C$, expressed by the ratio of the X-band VLBI flux density to RATAN-600 flux density.}
    \label{fig:flux_density}
\end{figure}
 
% \section{Methods, Observations, Simulations etc.}

\section{Methods} 

Our analysis of this high-redshift quasar sample employs a comprehensive methodological approach designed to extract maximum physical insight from the VLBI data while carefully accounting for observational limitations and potential biases. 
The key analysis steps are summarized below, with detailed procedures provided in Appendix \ref{app:data}.
% The analysis pipeline encompasses multiple stages, from initial data processing through detailed physical parameter estimation.

The VLBI data were obtained from the Astrogeo archive, which provides initial calibration for phase and complex gains. For about 12\% of observations requiring additional calibration (primarily due to complex source structures or technical issues), we implemented enhanced calibration using the DIFMAP software package \citep{1997ASPC..125...77S}. Quality control criteria included verification of amplitude and phase calibration, assessment of uv-coverage and sensitivity limits, and cross-validation with independent observations where available.

Source structures were characterized through model fitting using circular Gaussian components, a choice that provides good approximation for compact radio structures while minimizing free parameters. For multi-epoch sources, kinematic analysis required careful component identification and tracking, considering multiple parameters: component flux densities, sizes, positions relative to the core, and spectral properties where available.

% To quantitatively characterize source structures, we employ model fitting using circular Gaussian components. This choice of model provides a good approximation for compact radio structures while minimizing the number of free parameters, crucial for robust comparison across epochs and frequencies. The model fitting process implements a chi-squared minimization algorithm that accounts for the complex visibility data rather than image-plane fitting, ensuring maximum fidelity to the observational data \citep{1999ASPC..180..335P}.

% For multi-epoch sources, our kinematic analysis requires careful component identification and tracking. We develop a robust cross-identification scheme that considers multiple parameters: component flux densities, sizes, positions relative to the core, and spectral properties where available. This comprehensive approach helps mitigate the challenges posed by varying resolution and sensitivity across epochs, particularly important given our wide frequency range and long temporal baseline.

Rest-frame brightness temperatures were calculated for fitted components using standard formulae, with careful attention to resolution limits. For unresolved components, we derived conservative lower limits using the maximum resolution of each observation. Jet parameters were estimated using equipartition brightness temperature as a reference, though we note this assumption may not hold in all cases particularly in the most compact regions of AGN jets and in their maximum brightness state (see Appendix \ref{app:data}.4 for discussion).

% A key aspect of our analysis involves calculating rest-frame brightness temperatures for fitted components. We employ the standard formula but with careful attention to resolution limits and their impact on size measurements. For unresolved components, we derive conservative lower limits using the maximum resolution of each observation, calculated following the formalism of Ref. \citep{2005astro.ph..3225L}. This approach ensures reliable brightness temperature estimates while properly accounting for observational constraints.

% The derivation of jet parameters requires careful consideration of relativistic effects and projection. We estimate Doppler factors using the equipartition brightness temperature of the core as a reference, following methods established by  Readhead \citep{1994ApJ...426...51R} but with modifications to account for the high redshifts in our sample. 
However, we note that while this method is straightforward, it relies on the assumption of equipartition between particle and magnetic field energy densities, which may not hold in all cases, particularly in the most compact regions of AGN jets and in their maximum brightness state. Therefore, these Doppler factor estimates should be considered indicative rather than definitive, and are used in conjunction with other observables to constrain the intrinsic jet parameters. 
% These estimates, combined with proper motion measurements where available, allow us to constrain the intrinsic jet parameters.

For sources with multi-frequency data, spectral index maps were constructed using matched-resolution images. This process required careful image alignment and consideration of both thermal noise and systematic calibration uncertainties. Through this systematic approach, we extracted reliable physical parameters while maintaining careful attention to uncertainties and potential systematic effects.

% For sources with multi-frequency data, we construct spectral index maps using matched-resolution images. This process requires careful image alignment using optically-thin jet components as references, followed by convolution to a common beam size determined by the lowest frequency observation. We implement an error analysis approach that accounts for both thermal noise and systematic calibration uncertainties in the spectral index calculations.

% Through this comprehensive methodological approach, we aim to extract reliable physical parameters while maintaining careful attention to uncertainties and potential systematic effects. The resulting measurements form the basis for our investigation of high-redshift radio jet properties and their evolution with cosmic time.  Detailed descriptions of each methodological step are given in Appendix \ref{app:data}.

\section{Results} \label{sec:results}

\subsection{Source Compactness}

Our VLBI imaging analysis reveals distinctive patterns in the morphological properties of high-redshift radio-loud AGN. 
To quantify source compactness, we calculated the ratio $C = S_{VLBI} / S_{RATAN}$, where $ S_{VLBI}$ represents the integrated flux density measured from VLBI observations and $S_{RATAN}$ denotes the flux density measured by RATAN-600. We focused on S band (2.3 GHz) and X band (7.6 and 8.4 GHz) for this analysis, as these frequencies had the highest number of source detections.  This parameter $C$ serves as a key diagnostic tool, with values approaching unity indicating highly compact sources where most emission originates from parsec scales. Lower $C$ values suggest the presence of extended emission on scales larger than those probed by VLBI.

To compute the $C$ parameter, we used several strategies to deal with the sparse VLBI and RATAN-600 data. We prioritized contemporaneous observations, interpolated flux densities when frequencies did not match, and applied variability corrections for non-contemporaneous data within 3 months. VLBI-only data served as lower limits for $C$. These methods allowed us to calculate consistent $C$ values across our sample, despite the data limitations.

Analysis of the compactness distribution (Figure \ref{fig:flux_density}) reveals a pronounced broad peak in our sample. The vast majority of sources cluster tightly in the range of $C = 0.8-1.1$, with a peak around $C \approx 0.9$, indicating that most of the emission originates from parsec-scale structures detectable by VLBI.  For the compactness distribution analysis shown in Figure \ref{fig:flux_density}, we used contemporaneous single-epoch measurements directly from the Astrogeo catalogue, where $C$ typically ranges from 0.6 to 1.4. The values of $C > 1$ may occasionally occur due to source variability between VLBI and RATAN-600 observation epochs, even after applying our 3-month contemporaneity criterion.

Of the 102 parent sample, 16 sources lacking VLBI detections and being excluded in this study would likely show lower $C$ values if they could be measured. Including these sources would extend the distribution toward lower compactness values, though the predominance of compact sources in our sample would likely remain significant.

The overwhelming majority of sources in our sample display remarkably compact structures, with 35\% appearing as unresolved cores and 59\% showing core-jet morphologies. 
% 15 exceed 0.8
% 51 in total
% so the ratio is 
The compactness parameter serves as an independent metric that corroborates our morphological classification scheme (see discussion in Section \ref{sec:morphology}). Sources classified as core-dominated are expected to exhibit $C$ values approaching unity, while those with significant extended emission should display lower $C$ values. Indeed, the majority of our sources (30\%) demonstrate high $C$ values ($> 0.8$), consistent with their classification as core or core-jet morphologies. This high degree of compactness aligns with expectations for a sample of high-redshift quasars, where extended emission is often resolved out in VLBI observations \citep{2008A&A...484L..39F, 2022ApJ...937...19Z}.

\begin{figure}
    \centering
    \includegraphics[width=0.8\textwidth]{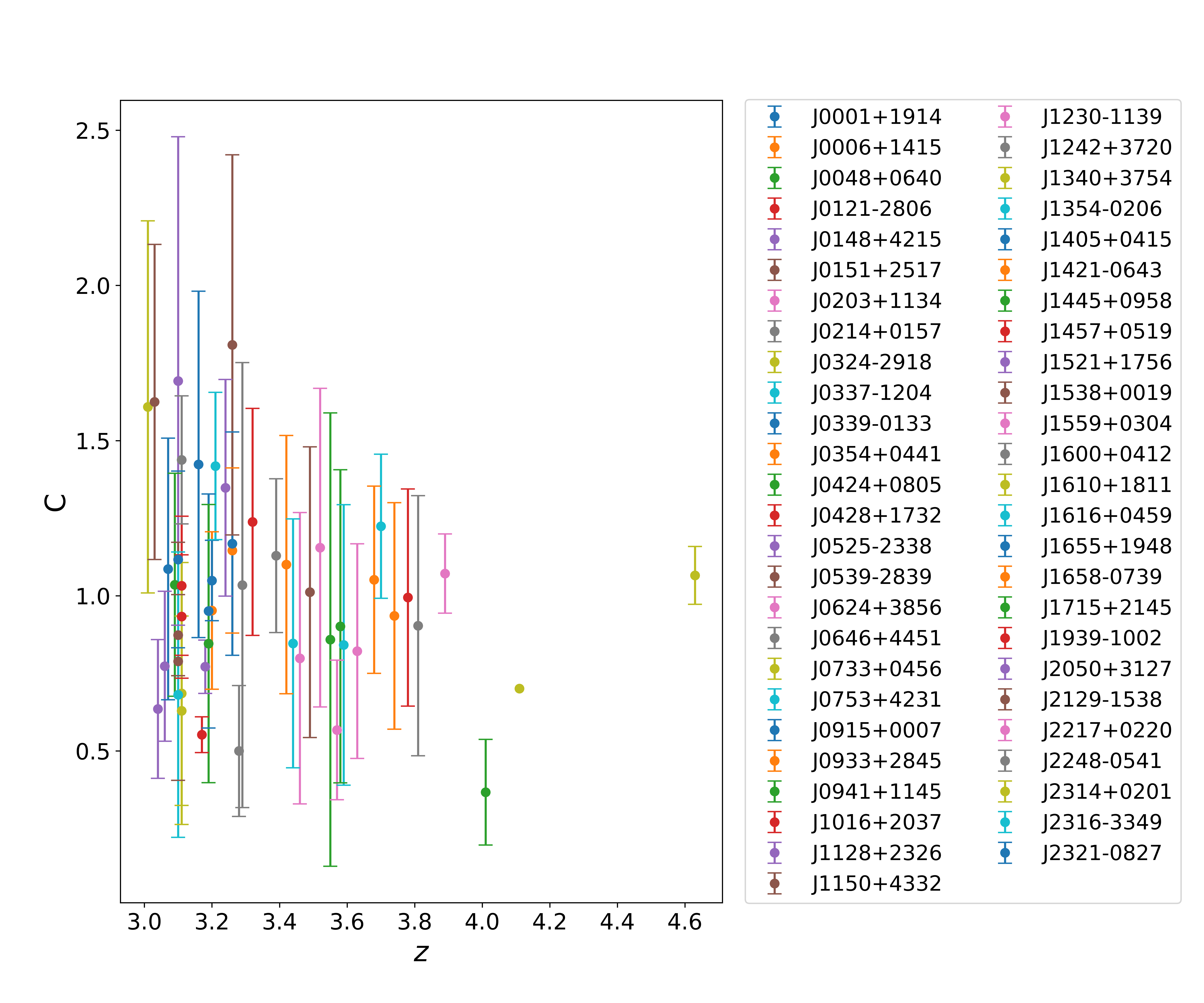}
    \caption{The mean value of compactness parameter of the quasars versus redshift $z$. The standard deviation is utilized for the error bar. In the case of a source with only one epoch measurement, the standard deviation is 0.}
    \label{fig:z_vs_compactness}
\end{figure}

To investigate evolutionary trends in our sample, we examined the relationship between compactness parameter $C$ and redshift (Figure \ref{fig:z_vs_compactness}). Each source in the plot shows a mean value of $C$ values with standard deviation due to variability across multiple epochs of observation. Across the redshift range ($z = 3.0-4.6$), the mean compactness values consistently cluster around $C \approx 1.0$, showing no significant systematic evolution with redshift. This stability suggests that the basic structural properties of radio-loud AGN cores remain relatively constant throughout this epoch of cosmic history.
Moreover, we observe a remarkable decrease in the dispersion of compactness value $C$ as redshift increases. While sources at $z \approx 3$ show a wide range of $C$ values (roughly 0.3 to 3.0), this spread progressively narrows at higher redshifts. This reduction in scatter could reflect either observational effects or physical processes. From an observational perspective, surface brightness dimming ($\propto (1+z)^{-4}$) might preferentially eliminate detection of extended structures at higher redshifts, effectively reducing the range of observable $C$ values. Additionally, our flux-limited sample may preferentially select more beamed, compact sources at higher redshifts, further reducing the observable range of structural variations.
The combination of stable mean compactness with decreasing variance provides important constraints for models of AGN evolution, though interpreting these trends requires careful consideration of selection effects and observational biases. Future observations with increased sensitivity to extended emission will be particularly valuable for understanding the reduced scatter at higher redshifts.

Only four sources (J0214+0157, J1457+0519, J1715+2145, J2314+0201, approximately 5\% of the sample) show notably lower $C$ values ($<0.7$), suggesting these are rare cases where substantial emission comes from larger-scale structures resolved by VLBI. This strong predominance of compact sources likely reflects both the selection effects of our flux-limited sample and the intrinsic nature of high-redshift radio-loud AGN, where extended emission may be suppressed by increased inverse Compton losses against the cosmic microwave background.

% \begin{figure}
%     \centering
%     \includegraphics[width=0.8\textwidth]{ratio/ratio_vs_Vs8.2.png}
%     \caption{The variability index V$_{S_{8.2}}$  versus the compactness parameter of the quasars.}
%     \label{fig:corr_compact_vari}
% \end{figure}

%TODO
% xlabel -> $C_max$
\begin{figure}
    \centering
    \includegraphics[width=0.8\textwidth]{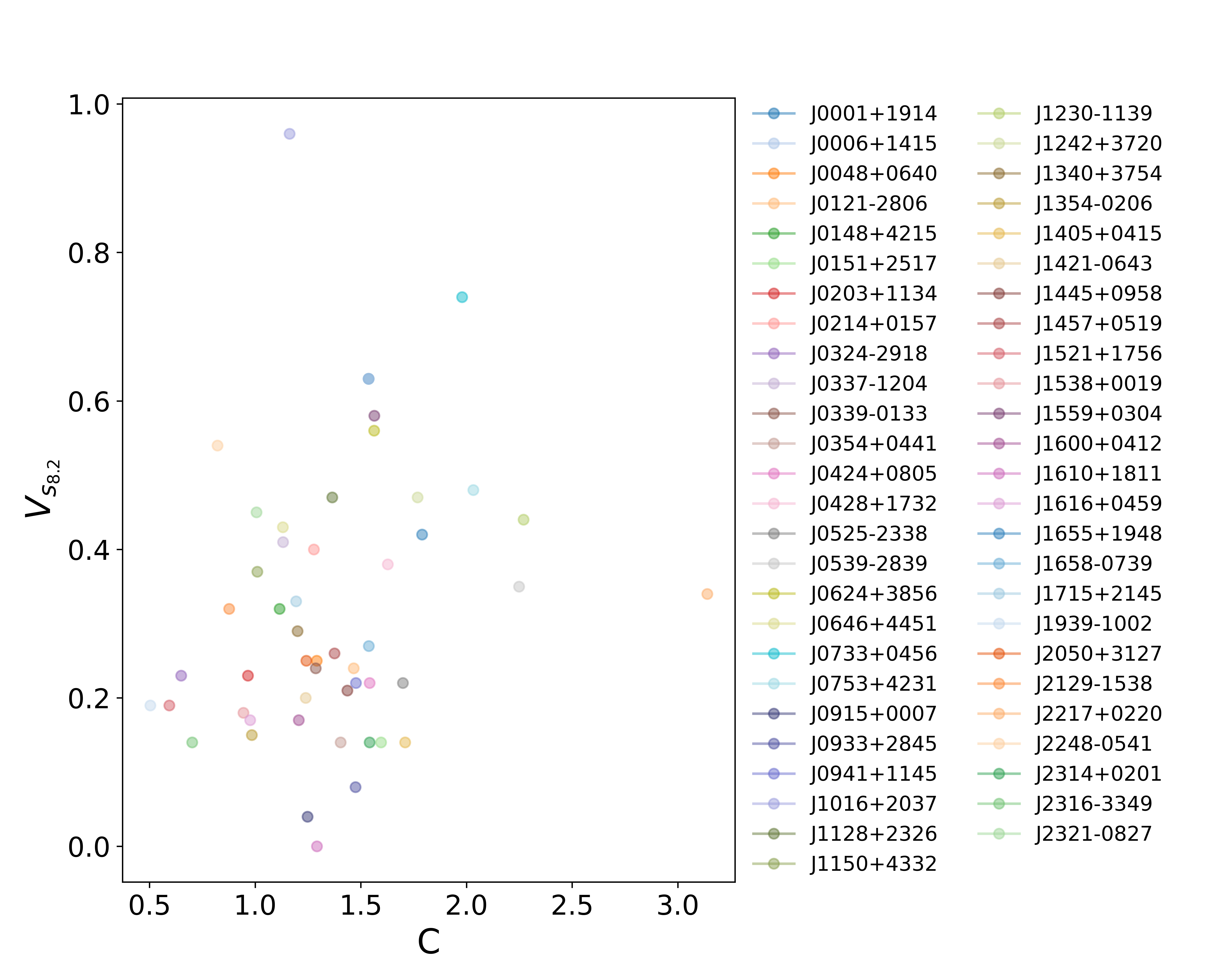}
    \caption{The variability index V$_{S_{8.2}}$  versus the maximum compactness parameter of the quasars.}
    \label{fig:corr_compact_vari_max}
\end{figure}

% \begin{figure}
%     \centering
%     \includegraphics[width=0.8\textwidth]{images/ratio_vs_z_max_Vs8.2.png}
%     \caption{The redshift $z$  versus the maximum compactness parameter of the quasars.}
%     \label{fig:z_vs_compactness}
% \end{figure}

% max corr is :  PearsonRResult(statistic=np.float64(0.2181974552283039), pvalue=np.float64(0.12399698913271848))

Figure \ref{fig:corr_compact_vari_max} illustrates the  relationship between the maximum source compactness ($C$) and variability index ($V_{S_{8.2}}$) for our sample. 
The variability index $V_{S_{8.2}}$  is calculated using the formula (1) in 
\citealt{2024Galax..12...25S} at X band.
Our correlation analysis yields a weak positive correlation coefficient of 0.22 between $V_{S_{8.2}}$ and $C$. While this correlation is not strong, it suggests a subtle tendency for more compact sources to display slightly higher variability amplitudes. However, the scatter in the distribution is considerable, with variability indices ranging from nearly 0 to about 0.95 across different $C$  values. Some of the most compact sources ($C > 1.5$) show moderate to high variability ($V_{S_{8.2}} > 0.4$), but there are also several compact sources with low variability levels. Different from the display in Figure \ref{fig:flux_density}, we plot the maximum value of $C$ observed across all available epochs for each source to capture the full range of structural variations. This multi-epoch approach results in some sources showing larger $C$ values (up to $\sim 3.0$) due to significant flux variability over the long monitoring period. These higher values typically correspond to epochs of enhanced source activity where the VLBI flux density temporarily exceeded the reference RATAN-600 measurements. This complex pattern suggests that while source compactness may influence variability behavior, it is likely not the dominant factor determining variability amplitude in high-redshift AGN. Other physical parameters, such as jet orientation, Doppler boosting, and intrinsic source properties, may play equally important or even more significant roles in determining the observed variability characteristics.

% \begin{figure}
%     \centering
% \includegraphics[width=0.8\textwidth]{ratio/ratio_vs_z_Vs8.2.png}
%     \caption{The compactness parameter versus reshift of the quasars.}
%     \label{fig:corr_compact_z}
% \end{figure}

\subsection{Morphology} 
\label{sec:morphology}

The morphological classification of high-redshift AGN presents unique challenges due to their extreme compactness, with many sources appearing unresolved even at VLBI resolutions. This limitation necessitates a comprehensive approach that integrates multiple lines of evidence beyond spatial structure alone.

Our classification methodology combines four key diagnostic indicators: spatial morphology, spectral properties, variability characteristics, and brightness temperature measurements. This integrated approach helps overcome the limitations of traditional classification schemes that rely primarily on resolved structure. For example, while jet components typically exhibit steep-spectrum optically thin emission above 2.3 GHz (corresponding to $\nu_{\rm int} \geq 10$ GHz in the observer's system of $z \geq 3$ sources), we find that even core components in our sample can display steep spectra between 5-8.4 GHz, likely due to the prevalence of peaked spectral distributions at high redshifts \citep{2021MNRAS.508.2798S}, which may indicate either youthful AGN activity or dense ambient medium confinement; in these cases, spectral indices measured between 5 and 8.4 GHz often fall in the steep-spectrum part of these peaked distributions.

To better understand these spectral characteristics, we constructed detailed spectral index maps for each source (Appendix \ref{sec:app:indexmap}, example in Fig. \ref{fig:vlbi:indexmap}), providing crucial additional constraints for morphological classification. Core components in our sample typically display high brightness temperatures ($10^{10}$ K), flat or inverted spectra, significant variability, and compact structure. In contrast, jet components generally exhibit lower brightness temperatures, steeper spectra, and more extended morphologies when resolved.

This multi-parameter classification framework allows us to confidently categorize sources even when traditional structural analysis proves insufficient, providing a robust foundation for investigating the physical properties of high-redshift radio-loud AGN.

\begin{figure*}
    \centering
    \includegraphics[width=0.4\textwidth]{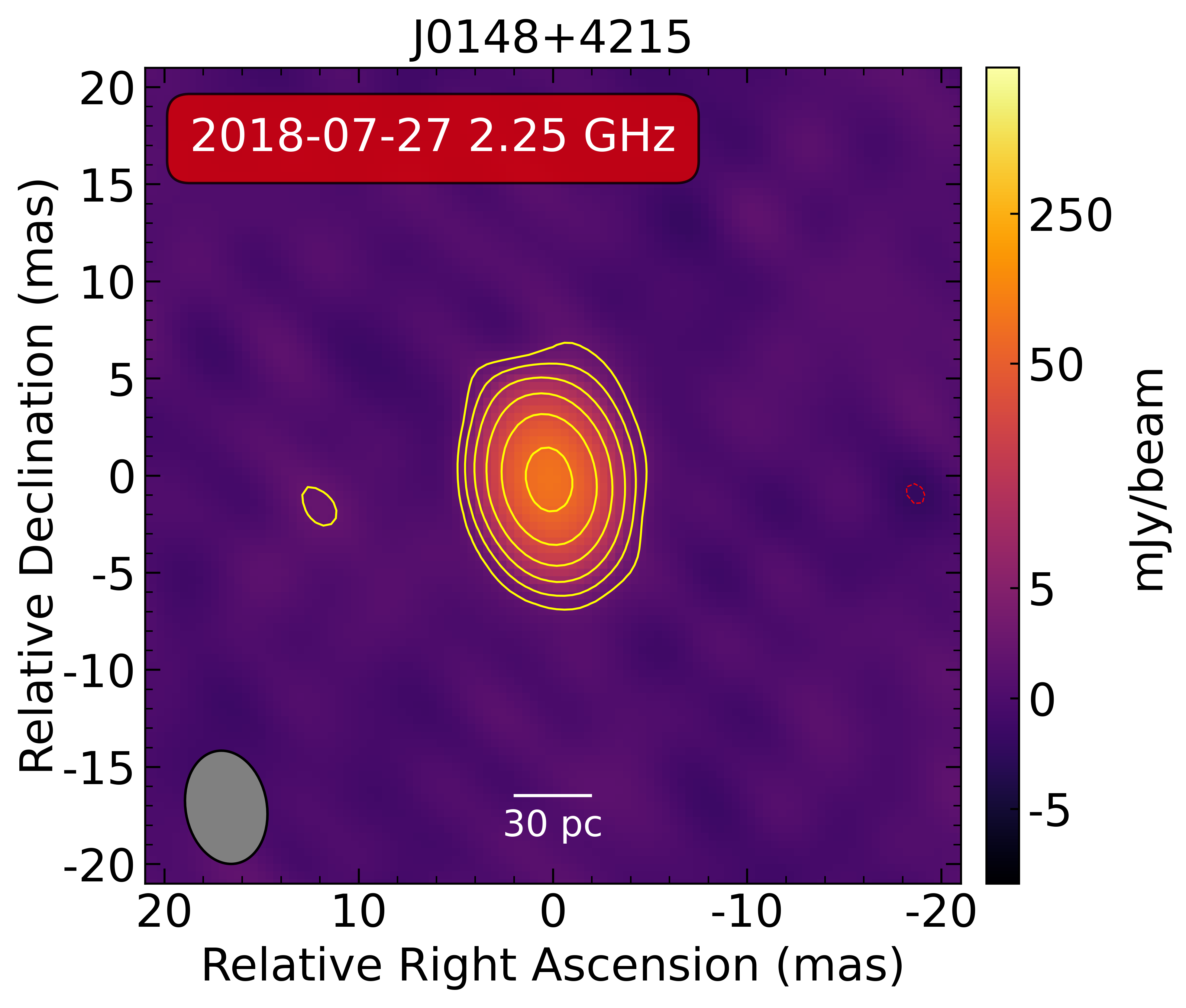}
    \includegraphics[width=0.4\textwidth]{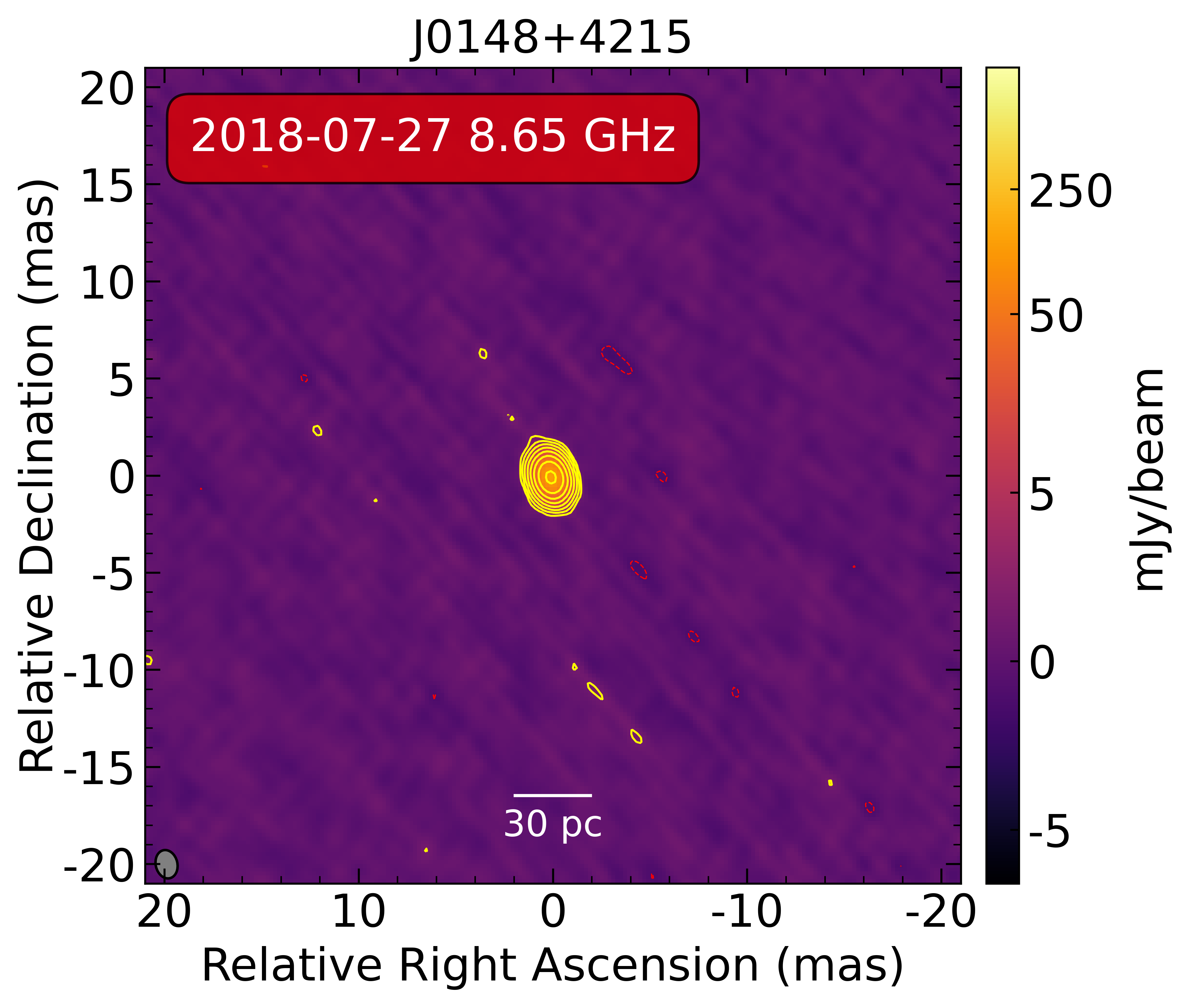}\\
    \includegraphics[width=0.4\textwidth]{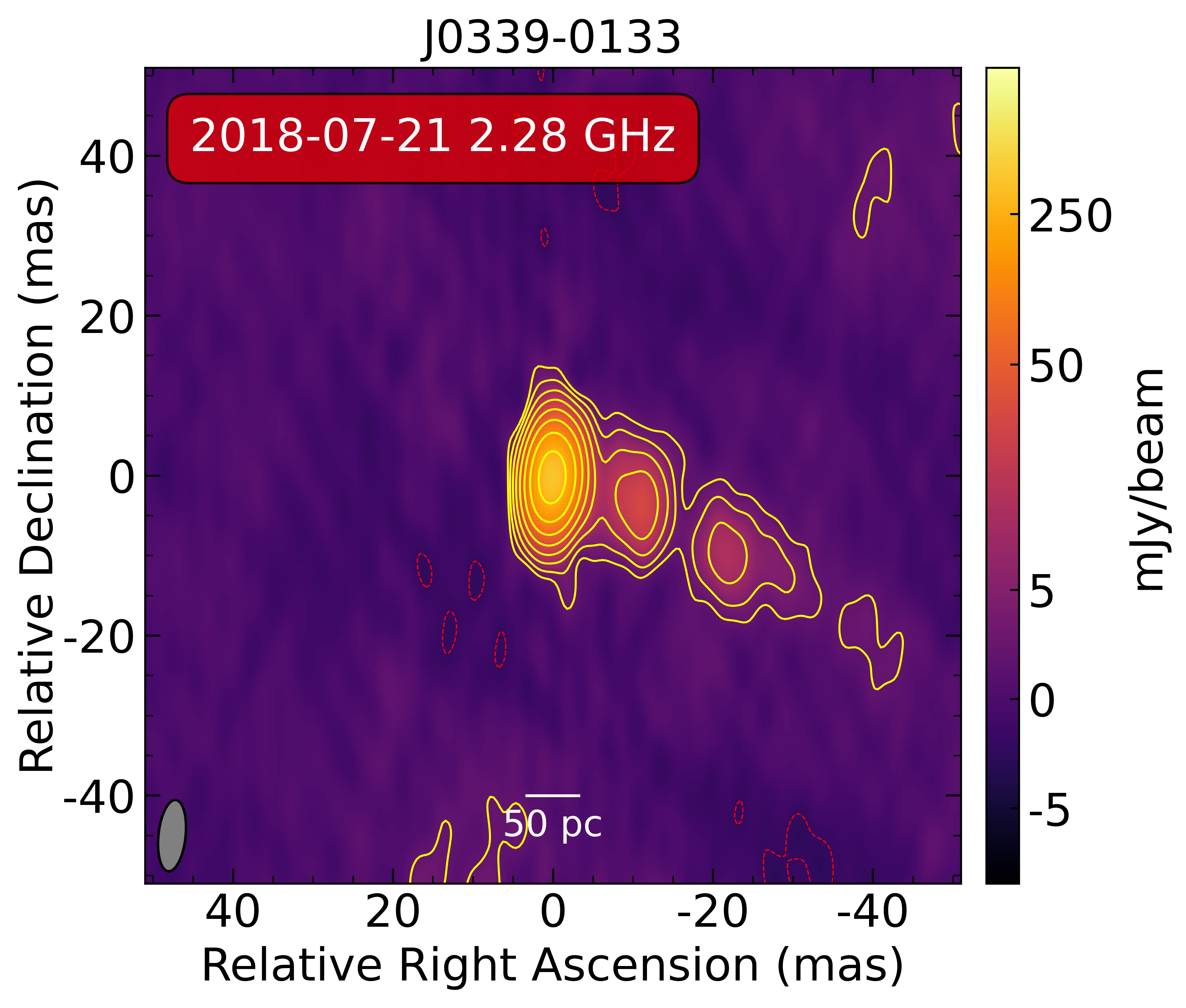}  
    \includegraphics[width=0.4\textwidth]{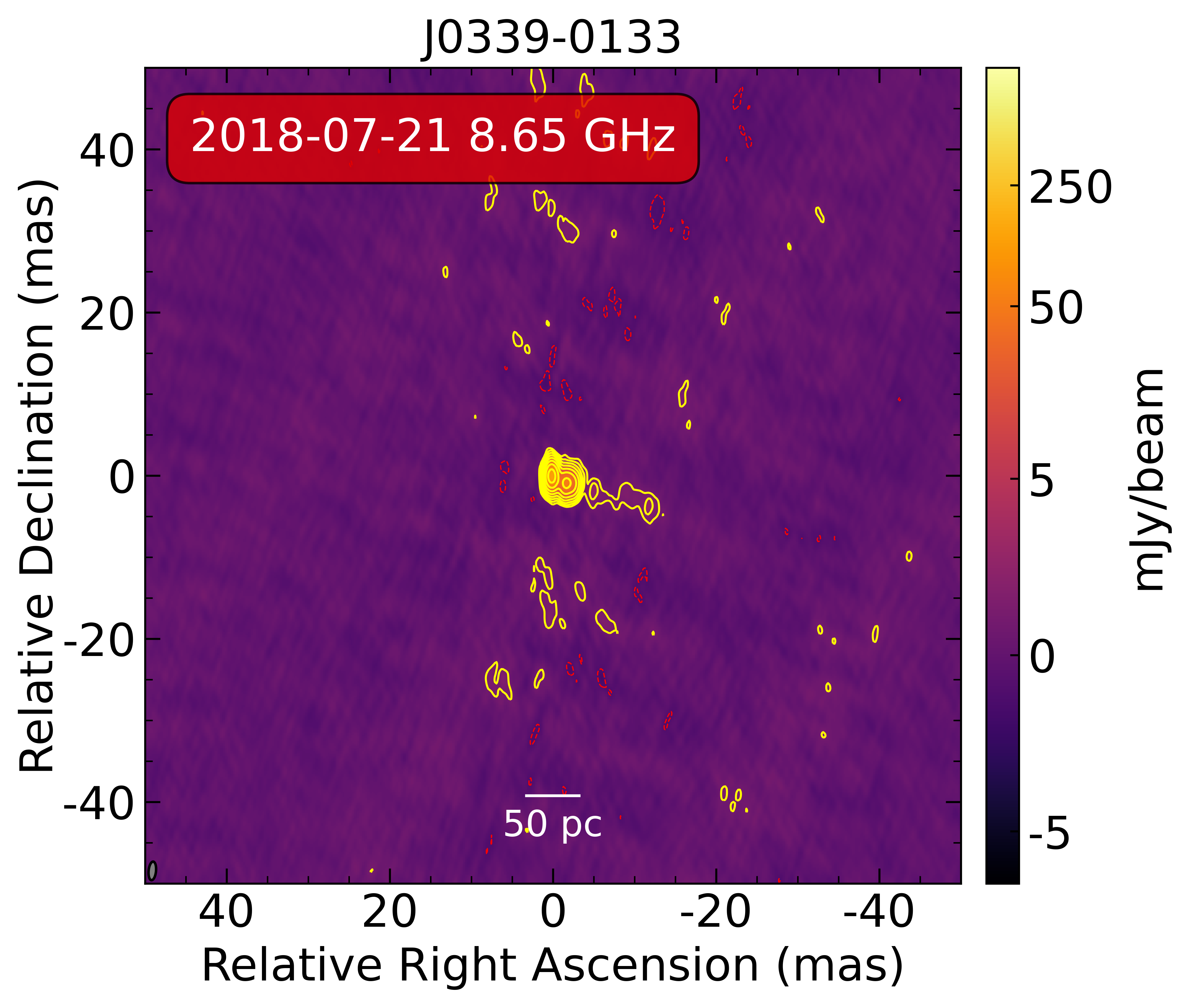} 
    \includegraphics[width=0.4\textwidth]{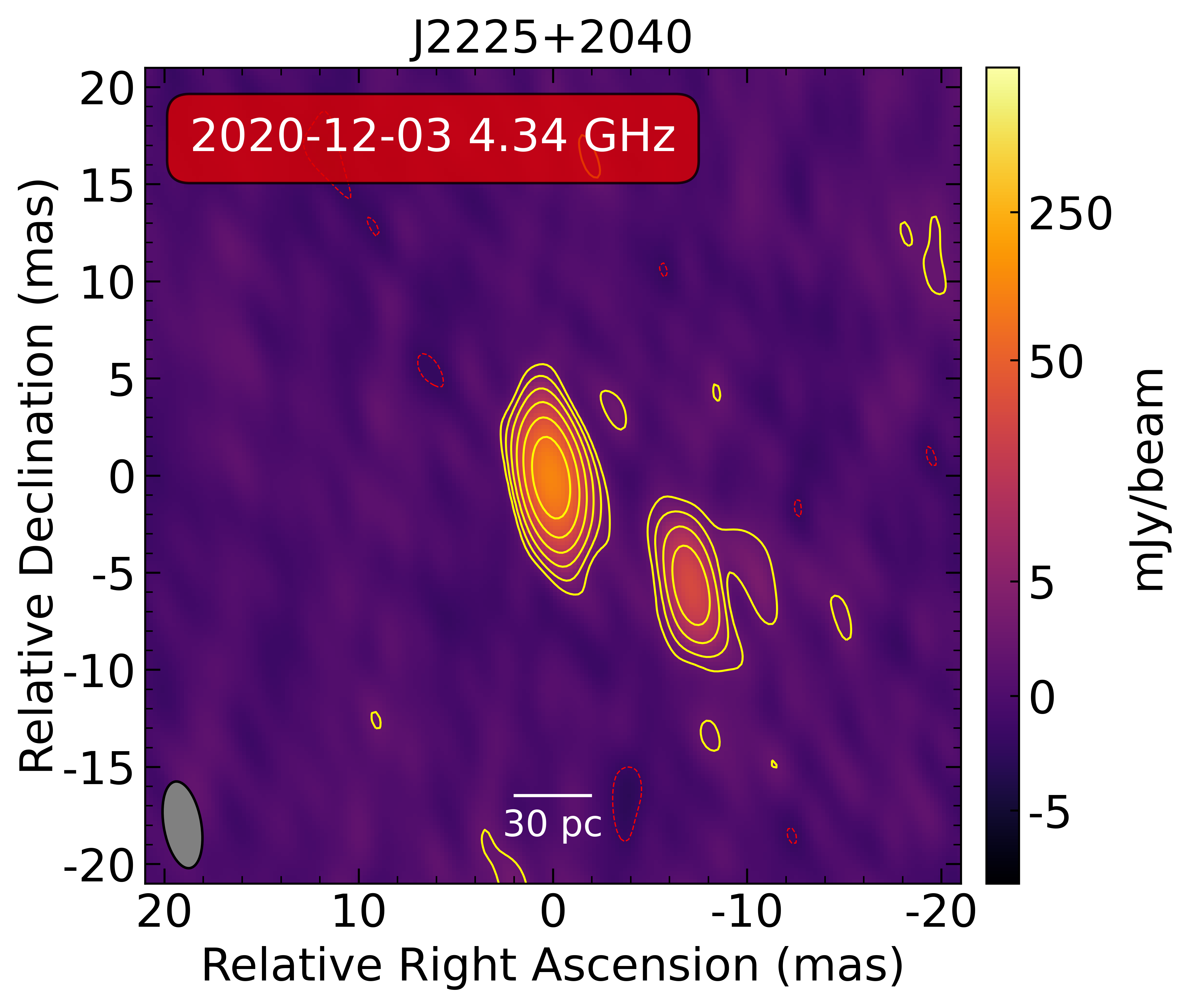} 
    \includegraphics[width=0.4\textwidth]{
    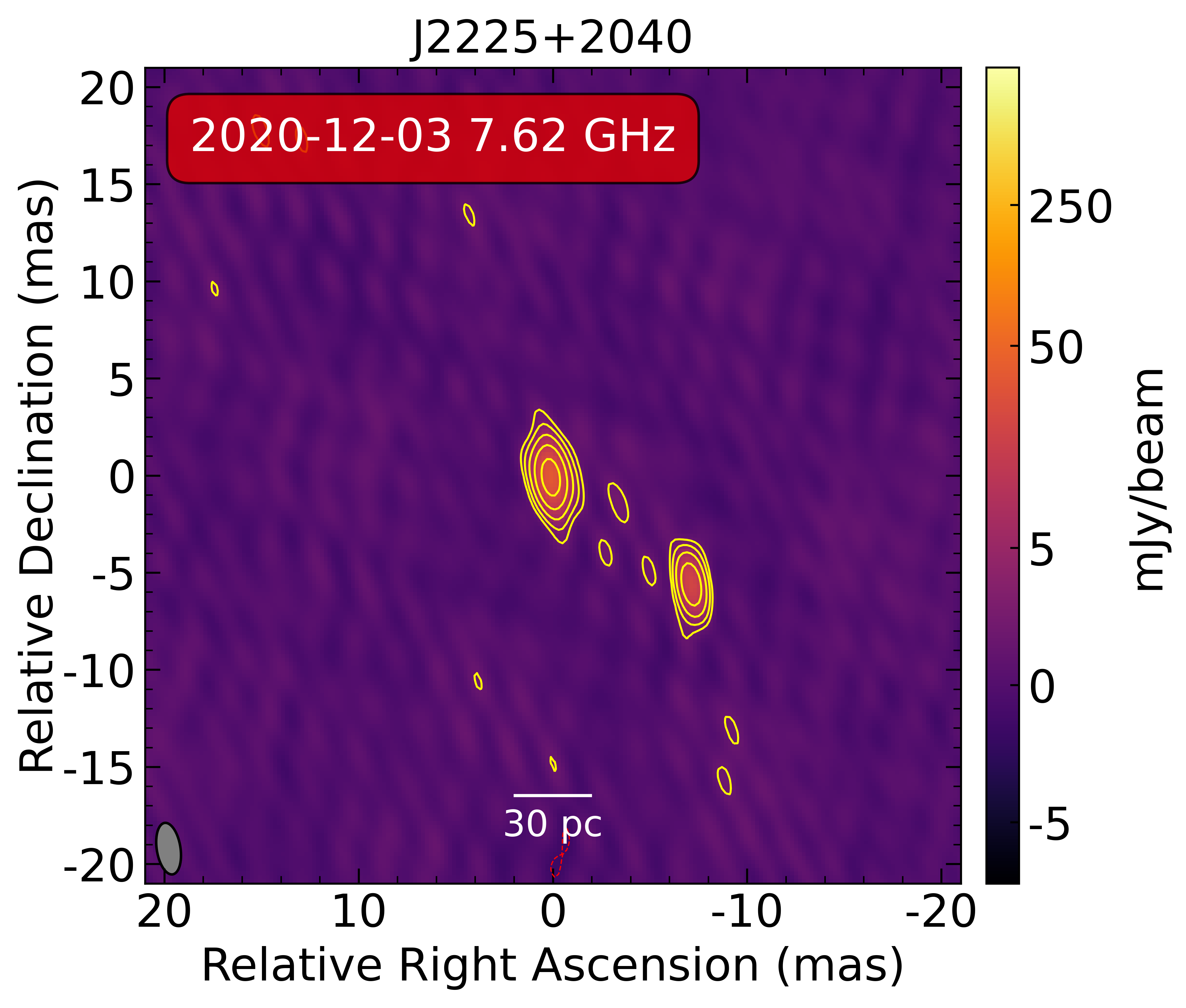} \\
    \caption{Example of VLBI images: compact core (top), core-jet (middle), core and double jets (bottom) complex morphology.}
    \label{fig:vlbi}
\end{figure*}

Based on these criteria, we classified sources into the following categories: compact core-only sources (C), core-jet structures (CJ), and compact double jets (CD). 
Figure \ref{fig:vlbi} shows the example images of these types. Class C sources remain point-like morphology at all resolutions, identified by its flat or inverted spectrum, high-level variability and high brightness temperature. Class CJ sources have a distinct core and one-sided weaker jet structure, typically showing a steep spectrum and brightness gradient along the jet axis. Class CD sources show two steep spectrum features with comparable brightness, similar to CSOs. 
For sources with multi-epoch observations, we ensured the classification was consistent across epochs, noting any significant structural changes.

% 8.973256125047943 \pm  2.1587323537677507
The core-only sources (Class C, accounting for 35\%) are characterized by a single, compact component that remains unresolved even at the highest available angular resolution ($\sim0.5$ mas at 8.4 GHz). These sources exhibit median brightness temperatures of $T_b = (8.97 \pm 2.15) × 10^{10} $K, approaching but generally not exceeding the inverse Compton limit. This suggests that while relativistic beaming is present, extreme Doppler boosting factors ($\delta > 10$) are rare in our high-redshift sample.
% Most of our sources (38\%) are unresolved at all observed frequencies, consistent with expectations for high-redshift, radio-loud quasars due to relativistic beaming of the core and cosmological surface brightness dimming of extended emission \citep[e.g.][]{1997A&A...318...11G, 2010A&A...524A..83F}. 
Some of the class C sources display flat spectra in RATAN-600 observations, suggesting they are likely blazars \citep{1999A&AS..139..545K}. A subset (55\%) show peaked spectra\citep{2021MNRAS.508.2798S}, potentially indicating young or recently reactivated AGN \citep{1998PASP..110..493O}.

Core-jet (CJ) sources comprise 59\% of our sample.
% display a compact, flat-spectrum core with an extended, steep-spectrum jet component. 
Their total flux density spectra range from flat to peaked, reflecting the complex interplay between core and jet emission at different frequencies. This spectral diversity highlights the importance of high-resolution VLBI observations in disentangling emission components in high-redshift quasars \citep{2012A&A...544A..34P}.

Core-double jet (CD) sources, while rare in our sample (5\%), present intriguing cases for study. These sources typically show GPS-type peaked spectra in their total flux density measurements, suggesting they may represent young radio sources in the early stages of evolution \citep{2012ApJ...760...77A, 1998PASP..110..493O}. The small fraction of complex or two-sided sources in our sample presents an interesting puzzle. Their rarity could reflect either genuine evolution in radio source properties with cosmic time or observational biases against detecting extended emission at high redshifts. Extended emission could be potentially missed in our observations, suggesting that the true fraction of complex sources might be higher.

Our classification is unavoidably subject to resolution limitations, especially for the most distant sources. Some core-jet sources might be misclassified as compact cores due to these constraints. Future observations with enhanced sensitivity and resolution may reveal additional structural details, potentially leading to reclassification of some sources \citep{2015aska.confE.143P, 2019ChJSS..39..242A, 2020AdSpR..65..850A}.

A particularly intriguing finding is the presence of large jet bending ($>90 ^\circ $) in five sources in our sample (J0203+1134, J1230-1139, J1405+0415, J1445+0958, J2003-3251), accounting for approximately 6\% of our sample. This fraction appears slightly higher than that seen in lower-redshift samples \citep{2021ApJ...923...30L}. This comparison warrants careful consideration.  The large jet bending may indicate stronger interaction with a denser circumnuclear medium, more frequent merger-induced perturbations or enhanced pressure gradients in the early Universe. Our flux-limited sample is inherently biased toward the most powerful radio sources at high redshift, which may be less susceptible to environmental effects due to their higher jet powers. A more complete understanding of jet bending mechanisms across cosmic time would require expanding studies to include weaker radio AGN at high redshifts, where lower-power jets would be more sensitive to interaction with the host galaxy environment and ambient medium. Additionally, the denser environments expected in the early Universe could play a more significant role in jet deflection for these lower-power sources.

This morphological analysis provides insights into jet physics and source evolution in the early Universe, though several observational biases must be carefully considered. First, VLBI observations inherently tend to miss low surface brightness emission, particularly at high redshifts where surface brightness sensitivity is further reduced by cosmological dimming. Second, our flux-limited sample preferentially selects brighter sources, which are more likely to have jets aligned closer to our line of sight, resulting in compact nuclei that can outshine extended jet emission due to Doppler boosting.
The high median brightness temperature ($T_b = (8.97 \pm 2.15) \times 10^{10}$K) observed in our Class C sources exceeds equipartition values, suggesting Doppler boosting and thus orientation bias, which is a notable contrast to studies of sources at $z>6$ \citep{2011A&A...531L...5F}. Therefore, while the prevalence of compact structures in our sample could indicate either young sources or strong environmental confinement of jets at high redshifts, these interpretations must be weighed against the aforementioned selection and observational biases.
These findings, when properly accounting for observational effects, provide constraints for models of AGN evolution and jet formation in the early Universe, though the true distribution of source properties may be broader than what our flux-limited, VLBI-selected sample reveals.

% This morphological analysis provides crucial constraints on jet physics and source evolution in the early Universe. The prevalence of compact structures suggests either that many sources are relatively young or that environmental effects play a significant role in confining jet expansion at high redshifts. These findings set important constraints for models of AGN evolution and jet formation in the early Universe.

% there are 55 sources in total which have data between 90 days
% J2129-1538 appears 41 times
% J0539-2839 appears 20 times
% J0646+4451 appears 23 times
% J1354-0206 appears 12 times

\subsection{Jet proper motion}
\label{sec:pm}

\begin{figure*}
    \centering
    \includegraphics[width=0.65\textwidth]{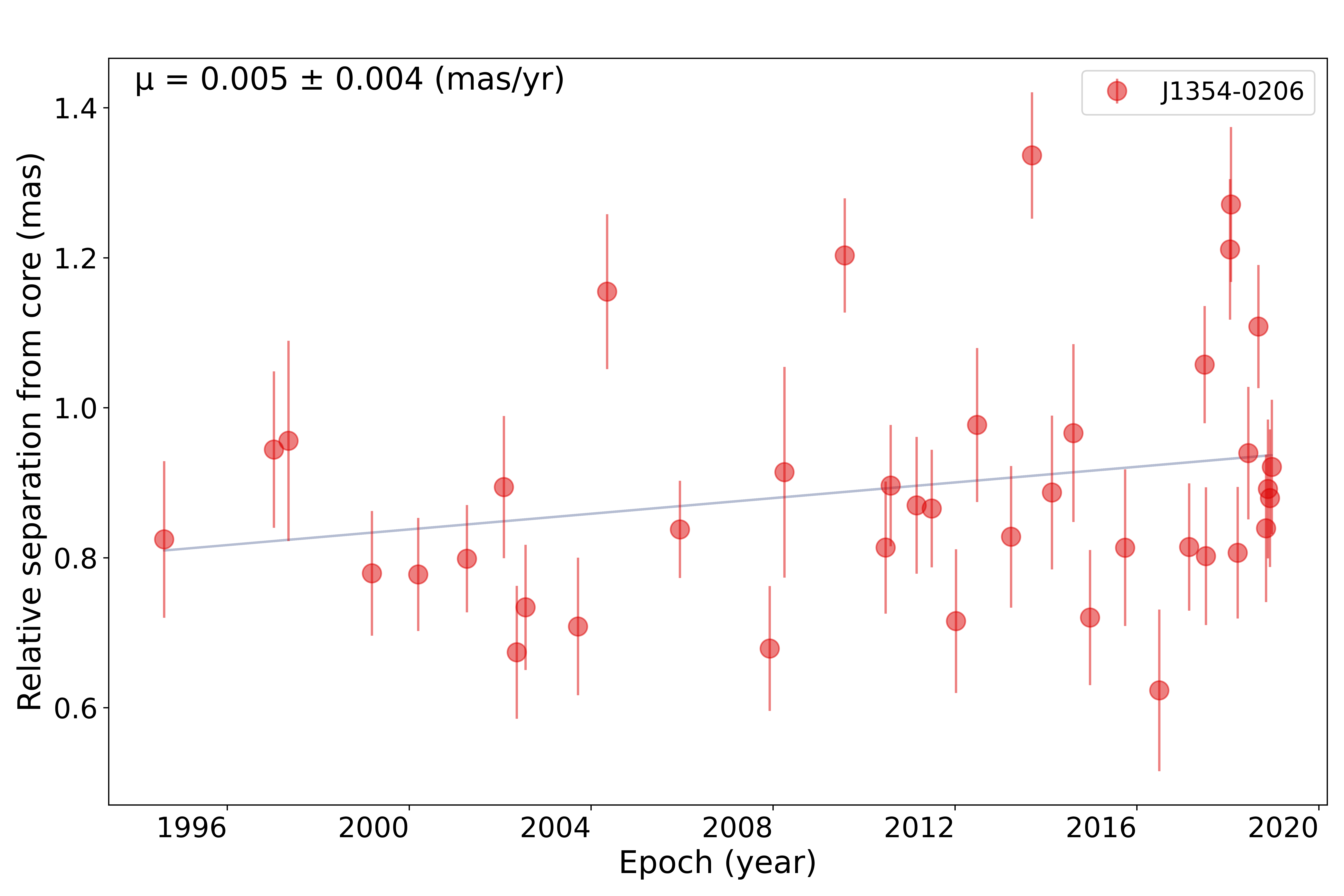}
    \includegraphics[width=0.65\textwidth]{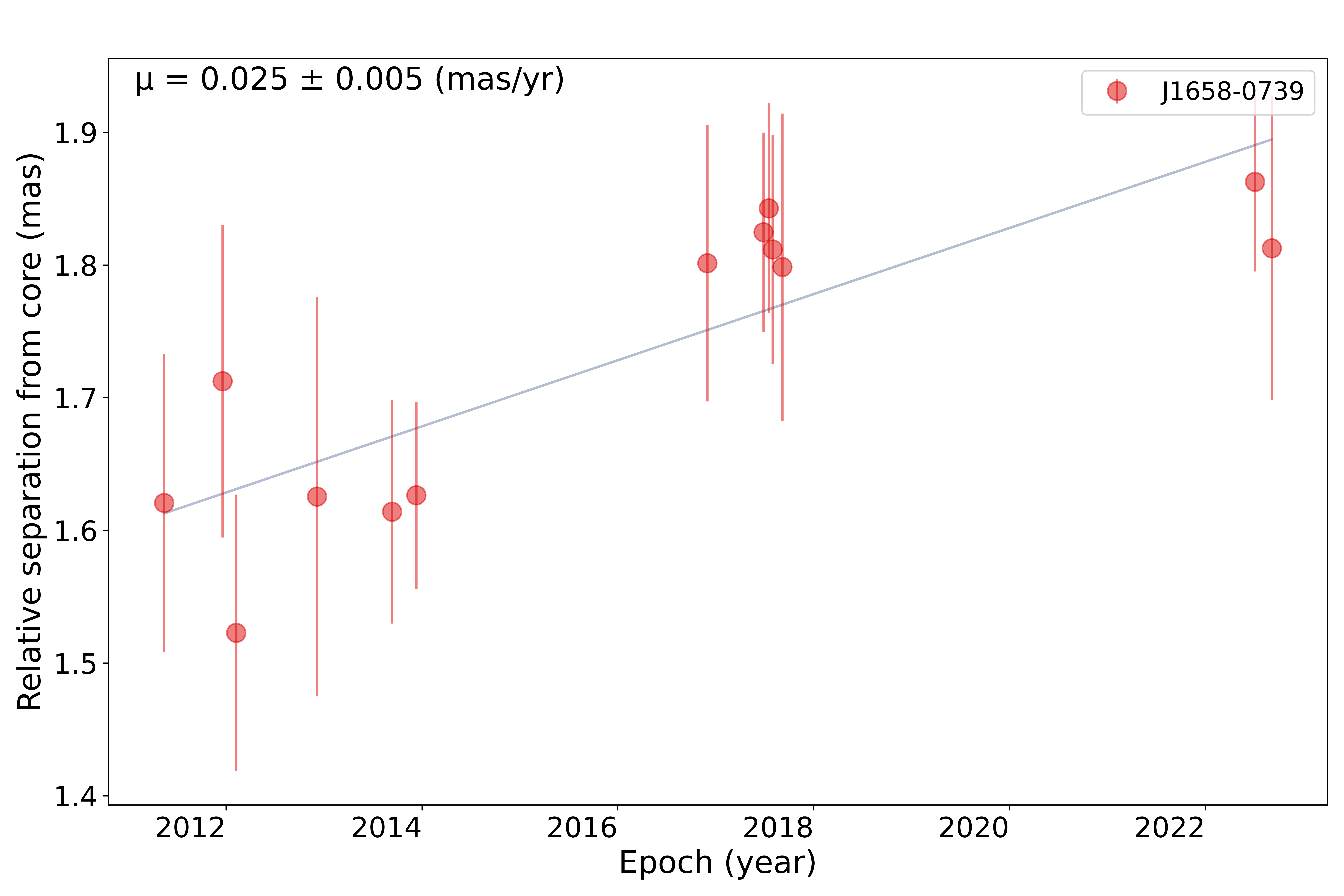}
    \caption{Examples of jet proper motions of the sample. Slow motion (J1354-0206) and moderate motion (J1658-0739). 
    % Corresponding J0203+1134, J0539-2839, J1354-0206 and J1658-0739 from top to bottom. 
    %All proper motion plots related to this study are available at \url{https://doi.org/10.5281/zenodo.14258042}.
    }
    \label{fig:vlbi:propermotion}
\end{figure*}

Our multi-epoch VLBI observations, spanning 5-15 years depending on the source, provide unprecedented insights into jet kinematics at high redshifts. 
Our kinematic analysis focuses on 57 sources with multiple epochs of X-band VLBI observations, providing optimal resolution and sensitivity for tracking jet component evolution. Within this subset, 34 sources have three or more epochs, enabling robust proper motion measurements through least-squares fitting of component trajectories. Examples of jet proper motions are shown in Figure~\ref{fig:vlbi:propermotion}. Other results can be accessed at Zenodo\footnote{\url{https://github.com/SHAO-SKA/Astro_AGN_catalogue.git}}. The distribution of jet proper motion is show in Figure~\ref{fig:vlbi:propermotion-all}.

For sources with only two epochs, we calculated jet proper motions directly from the positional changes of identifiable components.
The remaining sources with 3-9 epochs were analyzed using standard component tracking and linear fitting techniques.

Nine sources in our sample have more than 20 epochs of observations, allowing for detailed kinematic analysis. Some of these have been the subject of previous studies (J0339-0133\citep{2004ApJS..155...33S}, J0001+1914, J1658-0739\citep{2024MNRAS.530.4614K}, J1405+0415\citep{2018A&A...613A..74P}, J0203+1134, J0539-2839, J1354-0206, J2129-1538, and J0646+4451 \citep{2024A&A...685L..11G}). These well-sampled cases reveal complex kinematic behavior, with some components showing evidence for acceleration and/or non-linear trajectories. 
% Particularly noteworthy is J0646+4451, with 163 epochs spanning nearly three decades, which exhibits systematic acceleration in its inner jet region. 

% To investigate the kinematics of parsec-scale jets in our high-redshift sample, we conducted a detailed analysis of multi-epoch VLBI observations. Our dataset comprised 62 sources with two or more epochs of X-band (8.4 GHz) observations, providing an optimal combination of resolution and sensitivity for jet structure studies.

% For the 37 sources with three or more epochs, we employed a least-squares fitting approach to derive proper motions of distinct jet components. This method allows for a more robust determination of component velocities compared to simple two-point estimates, accounting for potential measurement uncertainties and intrinsic variations in component positions.
% J0941 0.09 mas/yr -> 10.01
% J0339 , 0.002 mas/yr -> 0.2 c

%The measured proper motions range from 0.01 to 0.48 mas yr$^{-1}$, corresponding to apparent superluminal speeds between $0.5\,c$ and $12.3\,c$ when converted to projected linear velocities. 
The measured proper motions range from 0.002 to 0.09 mas yr$^{-1}$, corresponding to apparent superluminal speeds between $0.2\,c$ and $10.0\,c$ when converted to projected linear velocities. 
The analysis reveals an intriguing distribution of jet speeds that challenge conventional expectations for high-redshift blazars (Figure \ref{fig:vlbi:propermotion-all}). 
This distribution of high-$z$ jets shows a marked difference from lower-redshift samples, where maximum apparent speeds often exceed $10\,c$. The lower apparent speeds in our high-redshift sample suggest either intrinsically lower jet bulk Lorentz factors or stronger environmental effects on jet propagation. 
% Most jet components exhibit apparent motions clustering around two distinct values: a population showing essentially stationary behavior ($\mu \leq 0.05$ mas yr$^{-1}$) and another displaying moderate superluminal speeds centered around 0.1-0.2 mas yr$^{-1}$. At a typical redshift of $z=3$, these speeds translate to apparent velocities of $\beta_{\rm app} \approx 0-2\,c$ and $8-15\,c$ respectively.

% A significant portion of these sources (XX) appear as unresolved cores even at X-band resolution, precluding their use in jet motion studies. For the remaining YY sources with resolved structures, we employed the following approach.

% details on speed.txt
%value < 2 : 48.28%
% 2 < value < 8 : 37.93%
% 2 < value < 8 : 13.79%

\begin{figure*}
    \centering
    \includegraphics[width=0.8\textwidth]{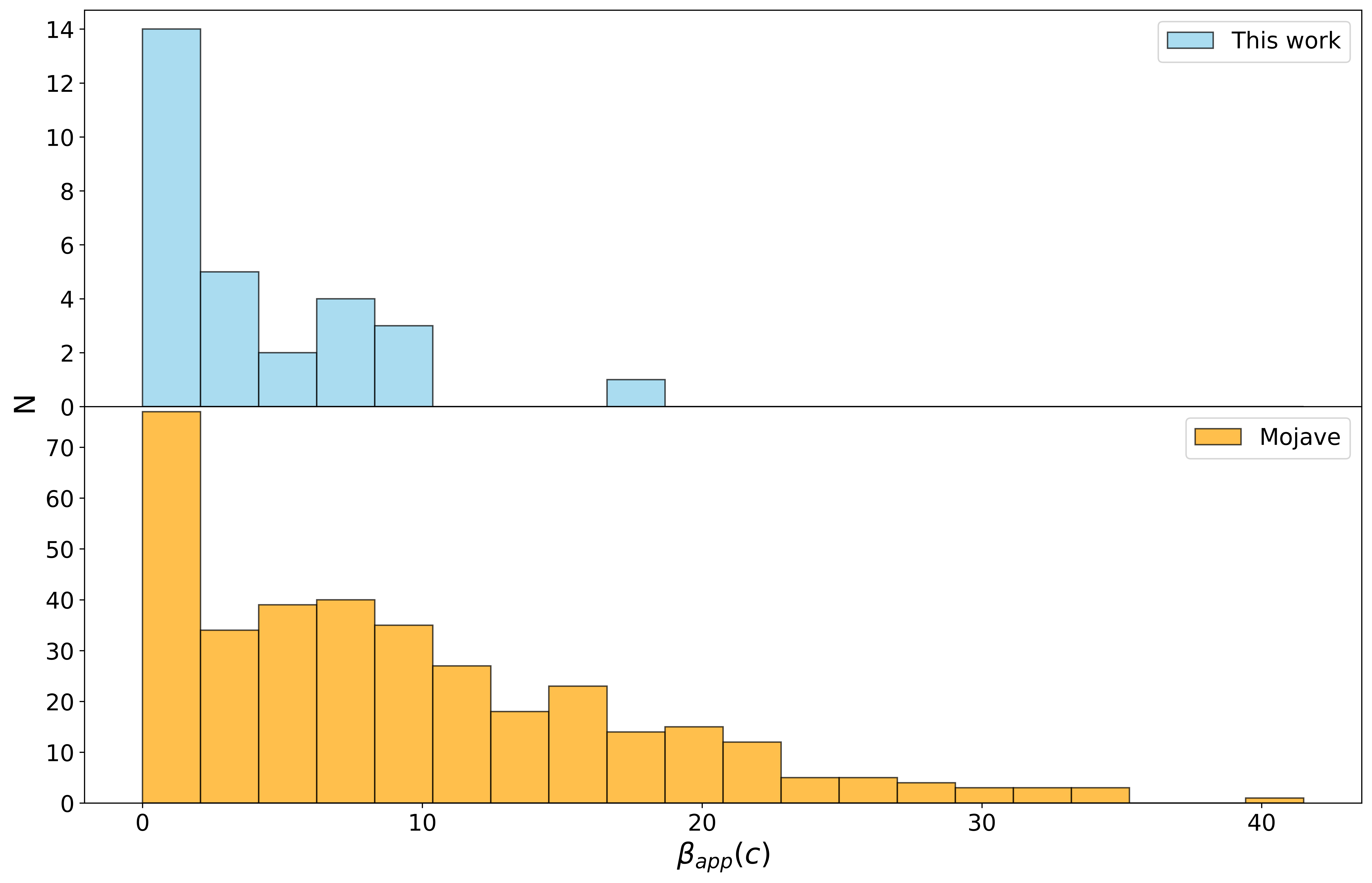}
    \caption{Histogram of maximum $\beta_{app}(c)$. Top: distribution of jet proper motions from this study; Bottom: distribution of maximum jet proper motions from MOJAVE programme (\citep{2016AJ....152...12L, 2021ApJ...923...30L}).
    }
    \label{fig:vlbi:speed}
\end{figure*}

Figure \ref{fig:vlbi:speed} presents a comparison between jet proper motions in our high-redshift sample and those from the MOJAVE sample (redshift range 0.05 to 3.40, median $z = 0.79$). While MOJAVE observations preferentially target sources with complex jet structures, potentially biasing their sample toward more dynamically active AGN, our comparison focuses on maximum jet speeds as these represent the highest achievable velocities in each population. This approach provides a conservative benchmark for comparing jet kinematics between epochs, as any selection bias toward more active sources in MOJAVE would only strengthen our finding that high-redshift jets show systematically lower speeds. The observed difference in velocity distributions persists even when accounting for these selection effects, suggesting a genuine evolution in jet kinematical properties with cosmic time.

% This bimodal distribution presents an intriguing puzzle.The observed jet speed distribution differs markedly from expectations based on lower-redshift samples, where apparent speeds exceeding $20\, c$ are common in blazar jets. 
The unexpected prevalence of slow or stationary components could be attributed to several possibilities: faster-moving components might exist on scales smaller than our resolution limit; sources viewed at larger angles to the line of sight will exhibit slower apparent motions; there may be a population of high-redshift sources with genuinely lower jet velocities.

The kinematic data also reveal interesting correlations with other source properties. Sources with lower apparent speeds tend to show higher fractional polarization and more stable flux densities, suggesting these may be viewed at larger angles to the line of sight despite their blazar-like core dominance. Conversely, the sources showing faster apparent motions typically display greater variability and lower polarization fractions, consistent with standard beaming models.

These findings have important implications for our understanding of jet physics in the early Universe. The scarcity of very high apparent speeds, combined with the trends in polarization and variability, suggests that either the intrinsic jet properties or their interaction with the environment differs significantly from what we observe in the local Universe. 
% This could reflect fundamental differences in jet formation and propagation mechanisms during earlier cosmic epochs.

% \subsection{Spectrum}

% use GLEAM, RACS, VLASS, RATAN600 

% use a fitting program FFA+SSA

\section{Discussion}

% 34.8% cs
% 59.3% cj
% cdj 6%
Our investigation of high-redshift radio-loud AGN using VLBI imaging reveals several notable characteristics that emphasize the profound influence of the early Universe environment on AGN jet evolution. Our VLBI imaging reveals a striking dominance of compact structures, with 94\% of sources showing either core-only or core-jet morphologies, significantly exceeding the fraction observed in lower-redshift samples, where more extended and complex structures typically dominate \citep{2017MNRAS.468.4992P}. While surface brightness dimming at high redshifts might bias against detecting large-scale structures \citep{1999A&A...342..378G}), our analysis indicates that observational effects alone cannot fully account for the pronounced morphological differences between high- and low-redshift AGN populations.

Proper motion measurements further stress fundamental distinctions in jet dynamics. We find that the apparent jet speeds in our high-redshift sample, ranging from 0–10 $c$, are systematically lower than those in well-studied low-redshift projects such as MOJAVE, where superluminal speeds frequently exceed 10 $c$ \citep{2021ApJ...923...30L}). This discrepancy persists even after accounting for flux-limited selection biases \citep{2021ApJ...923...67H}), and the absence of very fast jets ($\beta_{\rm app} > 10\,c$) \citep{2022ApJ...937...19Z} strongly suggests intrinsically lower bulk Lorentz factors or more pronounced environmental confinement at early epochs \citep{2020MNRAS.499..681M}). These findings align with theoretical expectations that the dense protocluster environments prevalent at high redshifts impose significant constraints on jet propagation \citep{2013ARA&A..51..105C, 2010A&ARv..18..279V}).

In addition to kinematic differences, the relationship between source compactness and variability offers valuable insights. Although only a weak positive correlation emerges between the compactness parameter $C$ and the maximum variability index $V_{s,max}$, this trend implies that more compact emission regions are associated with marginally greater variability amplitudes  \citep{1985ApJ...298..114M}). Yet, the substantial scatter in this correlation indicates that factors beyond emission region size ---such as jet composition, magnetic field configuration, or inhomogeneous external media --- also influence variability characteristics \citep{2018MNRAS.480.5517L}).

The prevalence of GPS and CSS sources within our sample further enriches our understanding of early AGN evolution. These spectral types may indicate either young AGN or ``frustrated'' sources confined by dense ambient medium  \citep{2021A&ARv..29....3O, 2012ApJ...760...77A}). The observed properties could be explained by either evolutionary effects or environmental conditions unique to the early Universe. At high redshifts, both scenarios are plausible: rapid supermassive black hole growth and abundant ambient gas could jointly govern jet evolution, hindering large-scale expansion and favoring early-stage or absorbed spectra \citep{2011MNRAS.417.2085V, 2012ARA&A..50..455F}.

Taken together, our findings strongly suggest that the early Universe exerts a more pronounced influence on AGN jet formation and evolution than previously recognized. Relative to lower-redshift systems, high-redshift jets display systematically lower speeds, more compact morphologies, and a higher fraction of peaked-spectrum sources, consistent with stronger environmental confinement and enhanced absorption processes \citep{2013MNRAS.430..174H, 2018A&ARv..26....4M} . Dense early environments may inhibit the development of large-scale radio structures and, through free-free absorption or synchrotron self-absorption, support a high occurrence of peaked-spectrum AGN \citep{2014MNRAS.438.2694G, 2016MNRAS.461L..21G}). Despite these constraints, the observed diversity in AGN properties suggests that intrinsic processes and line-of-sight effects continue to generate significant observed variability, even under challenging early-Universe conditions \citep{2014MNRAS.442L..81F}.

Comparisons with established low-redshift samples (e.g., MOJAVE)  \citep{2016AJ....152...12L, 2021ApJ...923...30L} confirm that the absence of ultra-fast jets and extremely high brightness temperatures in our high-redshift sources is not merely a selection artifact, but likely represents a genuinely different population  \citep{2022ApJ...937...19Z}). These results prompt a reevaluation of existing AGN evolutionary models, encouraging new theoretical frameworks that incorporate environmental density and cosmic epoch as key parameters \citep{2015MNRAS.446.2921F}. Furthermore, the systematically lower brightness temperatures in our sample suggest diminished Doppler boosting or altered microphysical conditions in the early Universe  \citep{2010A&ARv..18..279V, 2017A&ARv..25....2P}).

In addition, the high fraction of compact and peaked-spectrum AGN may indicate that we are observing many objects at nascent stages, either not yet evolved into large-scale radio sources or ``stalled" by dense media \citep{2012ApJ...760...77A, 2008MNRAS.388..625S}. This provides a unique window into the early growth and feedback of AGN in the formative stages of the Universe \citep{2012ARA&A..50..455F}. Our observations imply that the duty cycles, lifetimes, and evolutionary pathways of early AGN may differ significantly from their low-redshift counterparts, shedding new light on how supermassive black holes accumulated substantial mass so rapidly in the young cosmos \citep{2024MNRAS.528.5252M}.

Future VLBI observations with higher sensitivity and longer time baselines will be crucial to capturing subtle kinematic signatures and faint extended structures. Multi-wavelength observations, including millimeter/submillimeter and X-ray data, will help characterize the host galaxy environments, linking jet properties to their cosmic habitats \citep{2015aska.confE.173K, 2018Natur.553..473B}). 
%Future observations with next-generation facilities like the SKA and FAST Core Array and their VLBI components \citep{2015aska.confE.143P, 2024AstTI...1...84J}, combined with large-sample statistical studies, will test the trends identified here and refine AGN evolutionary models. Through expanded samples, improved sensitivity, and integrated theoretical efforts, we will achieve a more comprehensive understanding of jet physics, environmental effects, and the co-evolution of supermassive black holes and their host galaxies in the early Universe \citep{2017IJMPD..2630021M}.

\section{Conclusions} \label{sec:conclusions}

Our systematic VLBI study of 86 high-redshift ($z \geq 3$) radio-loud AGN has provided new insights into relativistic jets in the early Universe. Through detailed analysis of their morphology, kinematics, and spectral properties, we find evidence that jet physics at high redshifts may differ significantly from the local Universe.

The key findings from our study reveal three notable characteristics. First, the sources show a remarkably high degree of compactness, with VLBI-to-single-dish flux density ratios peaking sharply around 0.9. This compactness likely reflects both selection effects and enhanced inverse Compton losses against the cosmic microwave background at high redshifts. Second, the jet kinematics show predominantly low apparent speeds ($\beta_{\rm app} < 15\,c$), in contrast to lower-redshift samples where faster speeds are common for radio-brightest quasars. Third, the high fraction ($\sim$50\%) of peaked and steep-spectrum sources suggests a significant population of young or recently activated AGN and/or sources interacting with dense environments at high redshifts.

These results challenge our current understanding of AGN jet evolution. The combination of slower jet speeds, lower brightness temperatures, and high peaked-spectrum fraction points toward either stronger environmental effects or fundamentally different jet properties in the early Universe. The prevalence of young radio sources in our sample may provide crucial insights into the triggering mechanisms of AGN activity during this early cosmic epoch.

Future observations with next-generation facilities like the SKA and FAST Core Array and their VLBI components \citep{2015aska.confE.143P, 2024AstTI...1...84J}, combined with large-sample statistical studies, will be essential for expanding these studies to larger samples and higher sensitivity. 
These observations, combined with theoretical advances, will help distinguish between evolutionary and orientation effects, ultimately providing a clearer picture of how radio-loud AGN develop and evolve in the early Universe.

% By conducting the first systematic VLBI study of a substantial population of high-redshift, gigahertz-peaked-spectrum quasar candidates, this study can uniquely probe jet physics and SMBH growth in the early Universe. Resolving parsec-scale morphologies and brightness temperatures definitively discriminates between young GPS sources and blazars unresolved in single-dish data, providing direct views of compact symmetric objects tracing nascent jets. Measurements of the proper motions of jet components constrain the intrinsic jet velocities at early cosmological times. Elucidating the particle acceleration mechanisms, magnetic field configurations, and ambient medium interactions within these earliest collimated outflows shed light on jet formation paradigms in pristine environments. The definitive identification of the SMBHs power rare GPS quasars within the first billion years provides important clues to the rapid growth processes and accretion physics at work in the high-redshift Universe.

\section*{Acknowledgements}

This research is partly supported by the National SKA Program of China (2022SKA0120102, 2022SKA0130103, 2020SKA0110300). 
S.G. is supported by the Youth Innovation Promotion Association CAS Program under No. 2021258 and International Partnership Program of the Chinese Academy of Sciences under Grant No. 018GJHZ2024025GC.
T.A. acknowledge the support from the Xinjiang Tianchi Talent Program.
T.A. and Z.X. is supported by the FAST Special Program (NSFC 12041301).
Y.L. is funded by the National Natural Science Foundation of China (Grant No. 12403023), China Postdoctoral Science Foundation (certification number: 2023M733625, 2024T170968).
This work used resources from the China SKA Regional Centre prototype\citep{2022SCPMA..6529501A}.
%TODO CAS funding between CN and RU
Y.S. is supported by the Ministry of Science and Higher Education of the Russian Federation under the contract 075-15-2024-541.

%%%%%%%%%%%%%%%%%%%%%%%%%%%%%%%%%%%%%%%%%%
\vspace{6pt} 

%%%%%%%%%%%%%%%%%%%%%%%%%%%%%%%%%%%%%%%%%%
%% optional
%\supplementary{The following supporting information can be downloaded at:  \linksupplementary{s1}, Figure S1: title; Table S1: title; Video S1: title.}


%\begin{tabularx}{\textwidth}{lll}
%\toprule
%\textbf{Name} & \textbf{Type} & \textbf{Description} \\
%\midrule
%S1 & Python script (.py) & Script of python source code used in XX \\
%S2 & Text (.txt) & Script of modelling code used to make Figure X \\
%S3 & Text (.txt) & Raw data from experiment X \\
%S4 & Video (.mp4) & Video demonstrating the hardware in use \\
%\bottomrule
%\end{tabularx}
%}

%%%%%%%%%%%%%%%%%%%%%%%%%%%%%%%%%%%%%%%%%%
\authorcontributions{
Conceptualization, S.G. and T.A.; 
methodology, Y.L.; 
software, Z.X.; 
validation, C.L., S.G., Y.L., and Z.X.; 
resources, S.G., Y.S., T.M., A.W.; 
writing---original draft preparation, S.G., and T.A.; 
writing---review and editing, T.A.; 
visualization, S.G.; 
supervision, T.A.; 
project administration, S.G.. 
All authors have read and agreed to the published version of the manuscript.}

\dataavailability{
%RATAN data
All the data used in this study are publicly available and can be accessed through the Zenodo at \url{10.5281/zenodo.14258041} and the Github at \url{https://github.com/SHAO-SKA/Astro_AGN_catalogue}.
We acknowledge the use of archival calibrated VLBI data from the Astrogeo Center data base maintained by Leonid Petrov.
This research has made use of data from the MOJAVE database that is maintained by the MOJAVE team \citep{2018ApJS..234...12L}.
The National Radio Astronomy Observatory is a facility of the National Science Foundation operated under cooperative agreement by Associated Universities, Inc.
} 

% Only for journal Nursing Reports
%\publicinvolvement{Please describe how the public (patients, consumers, carers) were involved in the research. Consider reporting against the GRIPP2 (Guidance for Reporting Involvement of Patients and the Public) checklist. If the public were not involved in any aspect of the research add: ``No public involvement in any aspect of this research''.}

% Only for journal Nursing Reports
%\guidelinesstandards{Please add a statement indicating which reporting guideline was used when drafting the report. For example, ``This manuscript was drafted against the XXX (the full name of reporting guidelines and citation) for XXX (type of research) research''. A complete list of reporting guidelines can be accessed via the equator network: \url{https://www.equator-network.org/}.}

% Only for journal Nursing Reports
%\useofartificialintelligence{Please describe in detail any and all uses of artificial intelligence (AI) or AI-assisted tools used in the preparation of the manuscript. This may include, but is not limited to, language translation, language editing and grammar, or generating text. Alternatively, please state that “AI or AI-assisted tools were not used in drafting any aspect of this manuscript”.}

%\acknowledgments{In this section you can acknowledge any support given which is not covered by the author contribution or funding sections. This may include administrative and technical support, or donations in kind (e.g., materials used for experiments).}

\conflictsofinterest{The authors declare no conflicts of interest.} 

%%%%%%%%%%%%%%%%%%%%%%%%%%%%%%%%%%%%%%%%%%
%% Optional

%% Only for journal Encyclopedia
%\entrylink{The Link to this entry published on the encyclopedia platform.}

%%%%%%%%%%%%%%%%%%%%%%%%%%%%%%%%%%%%%%%%%%
%% Optional
\appendixtitles{yes} % Leave argument "no" if all appendix headings stay EMPTY (then no dot is printed after "Appendix A"). If the appendix sections contain a heading then change the argument to "yes".
\appendixstart
%\appendix
%\section[\appendixname~\thesection]{}
%\subsection[\appendixname~\thesubsection]{}

%\section[\appendixname~\thesection]{}
%All appendix sections must be cited in the main text. In the appendices, Figures, Tables, etc. should be labeled, starting with ``A''---e.g., Figure A1, Figure A2, etc.

%%%%%%%%%%%%%%%%%%%%%%%%%%%%%%%%%%%%%%%%%%
\begin{adjustwidth}{-\extralength}{0cm}

\reftitle{References}


%\bibliography{your_external_BibTeX_file}

%=====================================
% References, variant B: internal bibliography
%=====================================
\bibliography{universe-arxiv}

% If authors have biography, please use the format below
%\section*{Short Biography of Authors}
%\bio
%{\raisebox{-0.35cm}{\includegraphics[width=3.5cm,height=5.3cm,clip,keepaspectratio]{Definitions/author1.pdf}}}
%{\textbf{Firstname Lastname} Biography of first author}
%
%\bio
%{\raisebox{-0.35cm}{\includegraphics[width=3.5cm,height=5.3cm,clip,keepaspectratio]{Definitions/author2.jpg}}}
%{\textbf{Firstname Lastname} Biography of second author}

% For the MDPI journals use author-date citation, please follow the formatting guidelines on http://www.mdpi.com/authors/references
% To cite two works by the same author: \citeauthor{ref-journal-1a} (\citeyear{ref-journal-1a}, \citeyear{ref-journal-1b}). This produces: Whittaker (1967, 1975)
% To cite two works by the same author with specific pages: \citeauthor{ref-journal-3a} (\citeyear{ref-journal-3a}, p. 328; \citeyear{ref-journal-3b}, p.475). This produces: Wong (1999, p. 328; 2000, p. 475)

%%%%%%%%%%%%%%%%%%%%%%%%%%%%%%%%%%%%%%%%%%
%% for journal Sci
%\reviewreports{\\
%Reviewer 1 comments and authors’ response\\
%Reviewer 2 comments and authors’ response\\
%Reviewer 3 comments and authors’ response
%}
%%%%%%%%%%%%%%%%%%%%%%%%%%%%%%%%%%%%%%%%%%

%%%%%%%%%%%%%%%%% APPENDICES %%%%%%%%%%%%%%%%%%%%%

\appendix

\section{Sample Selection} \label{app:sample}

% 1213 total epoches
% 207 ecoches have nearest observation data (in 90 days)
\begin{landscape}
\begin{table}[h] 
\caption{The observational parameters of 1213 observation epochs \textsuperscript{1}. The columns are source name (1), redshifts (2), observation date (3), bandwidth (4), observation frequency (5), the synthesis beam FWHM and position angle (6, 7, 8), thermal noise level rms (9), integrated flux density, core flux density and core size from Gaussian model fitting (10, 11, 12), the ratio of core flux density versus integrated flux density (13), the VLBI integrated flux density versus RATAN flux density observed within 3 months (14).
\label{tab:sample}}
%\newcolumntype{C}{>{\centering\arraybackslash}X}
%\setlength{\tabcolsep}{6pt} % Adjust column spacing
%\renewcommand{\arraystretch}{1.2} % Adjust row spacing
%\begin{tabularx}{\textwidth}{CCCCCCCCCCCCCC}
\begin{tabularx}{\linewidth}{>{\raggedright\arraybackslash}X
>{\centering\arraybackslash}X
>{\centering\arraybackslash}X
>{\centering\arraybackslash}X
>{\centering\arraybackslash}X
>{\centering\arraybackslash}X
>{\centering\arraybackslash}X
>{\centering\arraybackslash}X
>{\centering\arraybackslash}X
>{\centering\arraybackslash}X
>{\centering\arraybackslash}X
>{\centering\arraybackslash}X
>{\centering\arraybackslash}X
>{\centering\arraybackslash}X
}
\toprule
name    & $z$  & epoch    & BW & $\nu$ & $\theta_{\text{maj}}$    &  $\theta_{\text{min}}$   &  P.A.     & rms   & $S_{\text{int}}$    & $S_{\text{core}}$ & Core Size   & $\frac{S_{\text{core}}}{S_{\text{int}}}$ & $\frac{S_{\text{VLBI}}}{S_{\text{RATAN}}}$   \\
(1) & (2) & (3) & (4) & (5) & (6) & (7) & (8) & (9) & (10) & (11) & (12) & (13) & (14) \\
 &  & & (MHz) & (MHz) & (mas)  & (mas) & ($^{\circ}$)& (Jy beam$^{-1}$)&(Jy)&(Jy)&(mas)&& \\ \midrule

J0001+1914 &		 3.10 &		 1996$-$01$-$02 &		 16.0 &		 8336.89 &		 2.93 &		 0.94 &		 104.1 &	1.15 &		 0.38 &		 0.38&		 0.16 &		 1.00 &		 0.99  \\
J0048+0640 &		 3.58 &		 2004$-$04$-$30 &		 32.0 &		 8646.22 &		 2.47 &		 1.08 &		 82.90 &	0.82 &		 0.07 &		 0.05&		 0.24 &		 0.74 &		 1.32   \\
J0121$-$2806 &		 3.11 &		 2005$-$05$-$12 &		 32.0 &		 8646.22 &		 3.51 &		 1.19 &		 101.8 &	1.15 &		 0.14 &		 0.14&		 0.80 &		 1.00 &		 0.87    \\
J0148+4215 &		 3.24 &		 2012$-$02$-$08 &		 32.0 &		 8642.24 &		 1.55 &		 1.28 &		 126.25 &	0.35 &		 0.12 &		 0.12&		 0.13 &		 1.00 &		 1.11    \\
J0151+2517 &		 3.10 &		 2005$-$07$-$09 &		 32.0 &		 8646.22 &		 4.20 &		 1.01 &		 105.7 &	1.02 &		 0.11 &		 0.11&		 1.66 &		 1.00 &		 1.07    \\
J0203+1134 &		 3.63 &		 1998$-$10$-$01 &		 16.0 &		 8644.23 &		 1.62 &		 0.73 &		 97.94 &	0.55 &		 0.53 &		 0.38&		 0.64 &		 0.72 &		 0.95   \\
J0257+4338 &		 4.06 &		 2005$-$06$-$30 &		 32.0 &		 8646.22 &		 1.96 &		 0.98 &		 83.80 &	0.50 &		 0.12 &		 0.10&		 0.30 &		 0.88 &		 0.88    \\
J0324$-$2918 &		 4.63 &		 2005$-$05$-$12 &		 32.0 &		 8646.22 &		 2.51 &		 0.95 &		 96.40 &	1.24 &		 0.09 &		 0.09&		 0.26 &		 1.00 &		 1.07    \\
J0337$-$1204 &		 3.44 &		 1997$-$05$-$07 &		 32.0 &		 8339.47 &		 2.80 &		 1.12 &		 88.75 &	1.89 &		 0.10 &		 0.10&		 0.18 &		 1.00 &		 0.45   \\
J0339$-$0133 &		 3.19 &		 2007$-$03$-$27 &		 32.0 &		 8642.24 &		 3.69 &		 2.57 &		 78.5 &		1.25 &		 0.13 &		 0.13&		 2.24 &		 1.00 &		 1.00    \\
J0354+0441 &		 3.26 &		 1995$-$07$-$15 &		 16.0 &		 8336.89 &		 2.34 &		 1.05 &		 92.44 &	1.24 &		 0.20 &		 0.2 &		 0.83&		 1.00 &		 0.93   \\
J0424+0805 &		 3.09 &		 2005$-$07$-$20 &		 32.0 &		 8646.22 &		 2.17 &		 0.91 &		 90.50 &	0.72 &		 0.33 &		 0.33&		 0.15 &		 1.00 &		 0.87   \\
J0428+1732 &		 3.32 &		 1996$-$01$-$02 &		 16.0 &		 8336.89 &		 2.67 &		 0.99 &		 100.75 &	1.26 &		 0.16 &		 0.16&		 0.31 &		 1.00 &		 1.03   \\
J0539$-$2839 &		 3.10 &		 1997$-$08$-$27 &		 32.0 &		 8339.47 &		 2.78 &		 0.95 &		 92.4 &		1.11 &		 0.87 &		 0.87&		 0.39 &		 1.00 &		 0.73    \\

\bottomrule
\end{tabularx}
\noindent{\footnotesize{\textsuperscript{1} The whole table can be found at \url{https://doi.org/10.5281/zenodo.14258041}.}}
\end{table}
\end{landscape}

\begin{landscape}
\begin{table}[h] 
\caption{Kinematics parameters of 34 high-$z$ quasars show resolved features from at least 3 epochs observations at X-band \textsuperscript{1}. The columns represent source name (1), brightness temperature $T_b$ in the unit of $10^{10}$K and the error (2, 3), Doppler factor $\beta$ (4), proper motion speed in the unit of light speed (c) and the error (5, 6), Lorentz factor $\Gamma$ (7), and viewing angle of the jet (8).  \label{tab:sample-pm}}
%\newcolumntype{C}{>{\centering\arraybackslash}X}
%\setlength{\tabcolsep}{6pt} % Adjust column spacing
%\renewcommand{\arraystretch}{1.2} % Adjust row spacing
%\begin{tabularx}{\textwidth}{CCCCCCCCCCCCCC}
\begin{tabularx}{\linewidth}{>{\raggedright\arraybackslash}X
>{\centering\arraybackslash}X
>{\centering\arraybackslash}X
>{\centering\arraybackslash}X
>{\centering\arraybackslash}X
>{\centering\arraybackslash}X
>{\centering\arraybackslash}X
>{\centering\arraybackslash}X
}
\toprule
 name & $T_b$ & $err_{T_{b}}$ & Doppler factor & Speed
& error Speed & Lorentz factor & Viewing angle  \\ 
(1) & (2) & (3) & (4) & (5) & (6) & (7) & (8)   \\
 & ($10^{10}K$) &   &  & (c) & &   & ($^{\circ}$)  \\
 \hline
\midrule

J0001+1914 & 102.08 & 0.31 & 20.42 & 1.71 & 1.43 & 10.30 & 0.46 \\
J0048+0640 & 6.38 & 0.11 & 1.28 & 1.94 & 1.03 & 2.50 & 41.44 \\
J0121-2806 & 1.43 & 0.01 & 0.29 & 6.96 & 5.50 & - & - \\
J0151+2517 & 0.26 & 0.01 & 0.05 & 0.79 & 2.17 & - & - \\
J0203+1134 & 6.91 & 0.01 & 1.38 & $\approx$ 0 & - & - & - \\
J0232+2317 & 19.39 & 0.14 & 3.88 & 8.97 & 0.56 & 12.42 & 10.75 \\
J0257+4338 & 9.74 & 0.05 & 1.95 & 0.99 & 0.22 & 1.48 & 27.61 \\
J0339-0133 & 0.18 & 0.01 & 0.04 & 0.20 & 1.00 & - & - \\
J0539-2839 & 41.37 & 0.05 & 8.27 & 5.03 & 1.10 & 5.72 & 6.18 \\
J0733+0456 & 34.33 & 0.15 & 6.87 & 4.51 & 3.05 & 4.98 & 7.72 \\
J0753+4231 & 11.68 & 0.06 & 2.34 & 0.53 & 0.97 & 1.44 & 12.69 \\
J0847+3831 & 26.16 & 0.23 & 5.23 & $\approx$ 0 & - & - & - \\
J0915+0007 & 4.54 & 0.06 & 0.91 & $\approx$ 0 & - & - & - \\
J0933+2845 & 0.34 & 0.01 & 0.07 & 7.91 & 10.86 & - & - \\
J0941+1145 & 1.25 & 0.01 & 0.25 & 10.01 & 0.93 & - & - \\
J1016+2037 & 5.06 & 0.02 & 1.01 & 3.78 & 1.42 & 8.05 & 27.84 \\
%J1128+23262 & 10.88 & 0.14 & 2.18 & -4.57 & 4.1 & 6.11 & -20.37 \\
J1230-1139 & 56.95 & 0.59 & 11.39 & 1.17 & 2.11 & 5.79 & 1.03 \\
J1242+3720 & 1.02 & 0.01 & 0.20 & 2.77 & 3.10 & - & - \\
J1340+3754 & 49.05 & 0.33 & 9.81 & $\approx$ 0 & - & - & - \\
J1354-0206 & 15.24 & 0.03 & 3.05 & 0.59 & 0.40 & 1.74 & 7.80 \\
J1356-1101 & 1.52 & 0.02 & 0.3 & 1.87 & 4.05 & - & - \\
J1405+0415 & 8.67 & 0.01 & 1.73 & 9.71 & 0.57 & 28.36 & 11.39 \\
J1421-0643 & 8.26 & 0.07 & 1.65 & 7.85 & 0.67 & 19.78 & 13.91 \\
J1430+4204 & 44.13 & 0.16 & 8.83 & 1.67 & 0.16 & 4.62 & 2.39 \\
J1445+0958 & 0.94 & 0.01 & 0.19 & 3.75 & 9.88 & - & - \\
J1521+1756 & 2.02 & 0.04 & 0.4 & 0.62 & 0.29 & - & - \\
J1538+0019 & 2.13 & 0.01 & 0.43 & 6.30 & 2.49 & - & - \\
J1658-0739 & 13.32 & 0.01 & 2.66 & 2.84 & 0.58 & 3.03 & 21.84 \\

\bottomrule
\end{tabularx}
\noindent{\footnotesize{\textsuperscript{1} The whole table can be found at \url{https://doi.org/10.5281/zenodo.14258041}.}}
\end{table}
\end{landscape}

%\begin{figure}[H]
%    \centering
%    \includegraphics[width=0.6\textwidth]{example.eps}
%    \caption{Radio morphology of high-$z$ quasars at mas scales.\label{fig:test}}
%\end{figure}
%\unskip

\section{Data analysis} \label{app:data}

\subsection{Self-calibration and imaging}

The VLBI data obtained from the Astrogeo archive have undergone initial calibration for phase and complex gains. For most sources, these pre-calibrated data were of sufficient quality to proceed directly to imaging. However, a subset of our data was affected by radio frequency interference (RFI), therefore the quality of the images created from the automatic imaging pipeline \footnote{\url{https://github.com/SHAO-SKA/vlbi-pipeline.git}} was not satisfactory and they necessitated additional processing steps. 
% In these cases, we performed manual self-calibration to obtain higher quality (lower noise level) images.

We implemented a hybrid approach combining automated and manual techniques to address the poor-calibrated data. We employed the {\sc Difmap} software package \citep{1997ASPC..125...77S} for both self-calibration and imaging. Our process began with the CLEAN algorithm in {\sc Difmap} to create an initial image, revealing the basic source structure. This step is crucial for identifying potential artifacts and guiding subsequent calibration efforts. We then performed iterative phase self-calibration, starting with longer solution intervals and gradually decreasing them. This approach effectively corrects for residual phase errors while minimizing the risk of introducing spurious features \citep{1999ASPC..180..187C}. Once phase solutions stabilized, we cautiously applied amplitude self-calibration, using longer solution intervals to preserve the integrity of the source structure. After the self-calibration process, we created the final images by using natural weighting. 

Throughout this process, we closely monitored the evolution of image quality and closure phases. This is essential to ensure that self-calibration improves data quality without introducing artifacts \citep{1978ApJ...223...25R}. Our final images typically achieved a dynamic range (ratio of peak flux to off-source rms noise) of several hundred to a few thousand, depending on source strength and data quality.

For sources with multiple epochs or frequencies, we applied consistent imaging parameters and self-calibration approaches. This consistency is crucial for our subsequent analysis of source variability and spectral properties, allowing for reliable cross-epoch and cross-frequency comparisons

\subsection{Model fitting}

To obtain a quantitative description of the radio structure, we conducted model fitting on the self-calibrated visibilities. We employed circular Gaussian models, which provide a good approximation for compact radio sources while minimizing the number of free parameters \citep{1999ASPC..180..335P}. The parameters of the fitted components, including peak flux density, integrated flux density, and deconvolved size, were then used to calculate each component's brightness temperature and spectral index (discussed in Section 4).

For sources with multi-epoch data, we analyzed the positional changes of VLBI components to calculate jet proper motion. This kinematic information, combined with the radio structure data, allows us to determine the geometry of the jet and estimate its kinematic age \citep{2009AJ....138.1874L}. The multi-epoch observations, spanning up to three decades in some cases, provide a unique opportunity to study the long-term evolution of jet structures in high-redshift quasars \citep{2015MNRAS.446.2921F, 2022ApJ...937...19Z}.

The angular resolution of the VLBI images varies depending on the observing frequency and array configuration, typically ranging from about 0.5 to 5 milliarcseconds. At the redshifts of our sources ($3.0 \leq z \leq 5.3$), this corresponds to linear scales of approximately 3 to 30 parsecs, allowing us to probe the inner jet zones where relativistic effects are most pronounced \citep{2020NatCo..11..143A}.

Component identification across multiple epochs is crucial for accurate proper motion analysis. We primarily focused on 8.4 GHz data, which offered the best combination of temporal coverage, resolution, and sensitivity for most sources. To identify and track components across epochs, we employed a multi-step approach: 
\begin{itemize}
    \item We compared the separation and position angle of each component relative to the core.
    \item We considered the evolution of flux density and size, expecting these properties to vary smoothly over time for a single physical component \citep{2019ApJ...874...43L}.
    \item In cases of ambiguous identification due to differences in resolution or sensitivity between epochs, we examined the overall structural evolution of the source and used physical arguments about plausible component motions to inform our decisions.
    \item When necessary, we referred to observations at other frequencies to aid in component identification.
\end{itemize}
This careful approach allowed us to track components reliably across epochs, even in the presence of varying image quality and resolution, ensuring the robustness of our kinematic analysis \citep{2012ApJS..198....5A}.

\subsection{Calculation of brightness temperatures}

The brightness temperature is a key parameter in characterizing the physical conditions in AGN jets and indicating the presence of relativistic beaming \citep{1988gera.book..563K}. For each fitted core component in our VLBI images, we calculated the rest-frame brightness temperature using the following formula \citep{2005AJ....130.2473K}:

\[ T_b = 1.22 \times 10^{12} (1+z) (S / \theta^2) \nu^{-2} [K] \]

where $z$ is the redshift of the source, $S$ is the flux density of the component in Jy, $\theta$ is the FWHM of the fitted circular Gaussian component in mas, and $\nu$ is the observing frequency in GHz.

For unresolved components, where the fitted size is smaller than the minimum resolvable size of the interferometer, we used the minimum resolvable size as an upper limit to calculate a lower limit on the brightness temperature. The minimum resolvable size was estimated using the formula 40 from \citet{2005astro.ph..3225L}.

We paid particular attention to the brightness temperature values of core components, as these can provide evidence for relativistic beaming when they exceed the inverse Compton limit of $\sim 10^{12}$ K for a stationary synchrotron source \citep{1969ApJ...155L..71K, 1981ARA&A..19..373K, 1994ApJ...426...51R}. Extremely high brightness temperatures ($T_b > 10^{13}$ K) may indicate either strong Doppler boosting or the presence of coherent emission mechanisms \citep{2007Ap&SS.311..231K, 2008ApJ...689..108L, 2009ApJ...703L.109S, 2016ApJ...820L...9K}.

In cases with multi-frequency observations of the same component, we used the brightness temperature measurements to estimate the spectral properties of the emission. A decrease in brightness temperature with increasing frequency can indicate synchrotron self-absorption, which is common in compact VLBI cores \citep{2020ApJS..247...57C}.

For sources with multiple epoch observations, we tracked the evolution of brightness temperature over time. Variability in brightness temperature can provide insights into changes in the physical conditions of the emitting regions, such as expansion, compression, or changes in the Doppler factor \citep{2005AJ....130.1418J, 2009A&A...494..527H}.

By analyzing the brightness temperatures across our sample of high-redshift quasars, we aim to characterize the emission properties of these distant AGN and compare them with lower-redshift populations to investigate any evolutionary trends in jet physics over cosmic time \citep{2015IAUS..313..327G}.

\subsection{Estimation of Doppler Factors}

To characterize the relativistic jets in our high-redshift quasar sample, we estimated the Doppler factors of the core components using a method based on the observed brightness temperature. This approach assumes equipartition between particle and magnetic field energy densities in the emission region \citep{1969ApJ...155L..71K, 1994ApJ...426...51R}.

The Doppler factor $\delta$ is estimated as: $\delta = T_{b,obs}/T_{b,eq} $, where $T_{b,obs}$ is the observed brightness temperature calculated as described in Section \ref{app:sample}, and $T_{b,eq}$ is the equipartition brightness temperature. Following \citet{1994ApJ...426...51R}, we adopt $T_{b,eq} = 5 \times 10^{10}$ K as the equipartition brightness temperature. This value represents the maximum brightness temperature attainable under conditions of equipartition between the energy densities of the radiating particles and the magnetic field.

Uncertainties in the Doppler factor estimates were calculated by propagating the uncertainties in the observed brightness temperature. For sources with multiple epoch observations, we calculated the Doppler factor at each epoch, allowing us to track potential variations over time. Such variations can provide insights into changes in the jet orientation or intrinsic properties.

While this method is straightforward and directly tied to our brightness temperature measurements, it relies on the assumption of equipartition, which may not hold in all cases, particularly in the most compact regions of AGN jets and in their maximum brightness state \citep{2006ApJ...642L.115H}. Therefore, we treat these Doppler factor estimates as indicative rather than definitive, and use them in conjunction with other observables to build a comprehensive picture of the jet properties.

By comparing our derived Doppler factors with those of lower-redshift samples \citep{2009A&A...494..527H, 2018ApJ...866..137L, 2020ApJS..247...57C}, we aim to investigate whether there are systematic differences in the beaming properties of high-redshift quasars that might indicate evolutionary trends in jet dynamics or orientation effects in flux-limited samples.

\subsection{Proper Motion Measurements for Multi-Epoch Sources}

% 34 sources have the proper motion results
For the 34 sources in our sample with $\geq 3$ epochs of VLBI observations, we performed proper motion analysis to study jet component kinematics. This analysis provides crucial information about jet speeds, which can be used to constrain the geometry and physical properties of the jets.

We identified and tracked individual components across epochs using a combination of visual inspection and quantitative criteria. The separation and position angle of the jet component with respect to the core component, which is assumed to be stationary, are calculated. Components were considered to be the same across epochs if their positions relative to the core were consistent with linear motion, their position angles, their flux densities and sizes evolved smoothly \citep{2019ApJ...874...43L}. 

Proper motions were calculated by fitting linear functions to the component positions over time. For a few sources, with multi-epoch data covering $>30$ year time baseline, we also tested for potential acceleration by fitting second-order polynomials to the component trajectories \citep{2009AJ....138.1874L}. Acceleration was considered significant if the improvement in the $\chi^2$ of the fit was greater than 99\% confidence level compared to the linear fit.
We converted angular speeds to apparent speeds in units of c using the formula: 
$\beta_{app} = \mu D_{A} / c(1+z)$
where $D_A$ is the angular size distance to the source and $z$ is its redshift \citep{2004ApJ...609..539K}.

%For sources with measured proper motions, we estimated the jet viewing angle θ and the bulk Lorentz factor Γ using the equations in \citet{1993ApJ...407...65G}.

% $\beta_{app} = \beta \sin \theta / (1-\beta \cos \theta)$,
% $\delta = [\Gamma (1-\beta \cos \theta)]^{-1}$
% where $\beta$ is the intrinsic jet speed and $\delta$ is the Doppler factor estimated in Section 3.4.

The uncertainties in proper motion measurements were estimated using least square method, 
taking into account the uncertainties in component positions and potential systematic errors in the VLBI astrometry \citep{2013AJ....146..120L}.

\begin{figure*}
    \centering
    \includegraphics[width=0.95\textwidth]{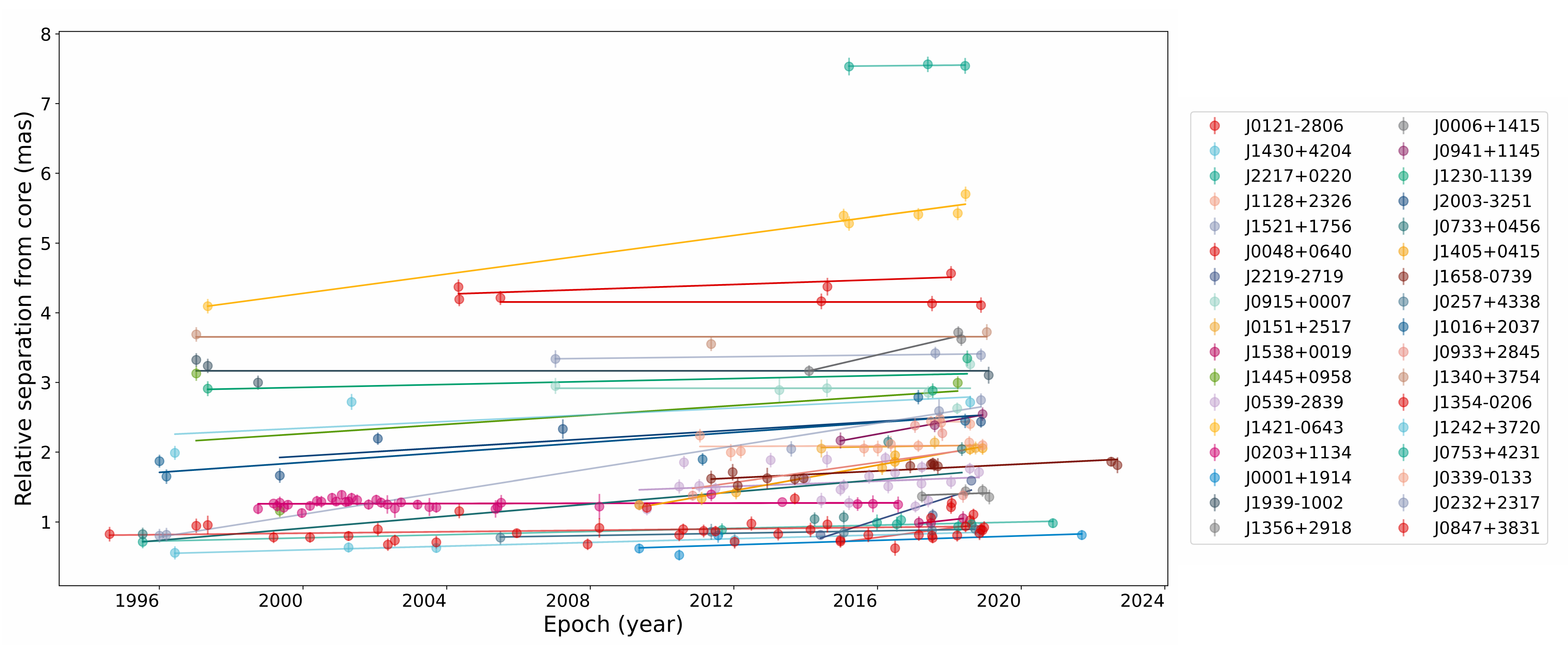}
    \caption{Distribution of jet proper motions in X band.}
    \label{fig:vlbi:propermotion-all}
\end{figure*}

\subsection{Spectral index map}
\label{sec:app:indexmap}

For sources with multi-frequency VLBI data, we created spectral index maps to further investigate the jet structure and emission mechanisms. Most sources in our sample have dual-frequency observations, either S/X (2.3/8.4 GHz) or C/X (4.7/7.6 GHz) band combinations. 

We focused our spectral index analysis on sources with parsec-scale extended jets, as these offer the most informative view of jet physics. Our approach was as follows:
for sources resolved in S band, we used S/X band images for spectral index mapping; 
for sources compact in S band but resolved in C and X bands, we produced C/X band spectral index maps;
we omitted spectral index mapping for sources appearing compact and unresolved at all available frequencies, as these would not provide meaningful spatial spectral information.

To create the spectral index maps, we followed these steps:

1. Image alignment: We aligned images at different frequencies using the 8-GHz VLBI core position as a reference point.

2. Image convolution: We convolved the higher-frequency images to match the resolution of the lowest-frequency image in each pair.

3. Spectral index calculation: We calculated the spectral index $\alpha$ ($S \propto \nu^\alpha$) for each pixel where the flux density exceeded 5 times the rms noise level in both images.

The uncertainty in the spectral index measurement comes from the uncertainty in the flux densities, which can be calculated by the following equation : 

\begin{equation}
\Delta\alpha = \frac{1}{\ln \frac{\nu_1}{\nu_2}} \sqrt{\left(\frac{\Delta S_1}{S_1}\right)^2 + \left(\frac{\Delta S_2}{S_2}\right)^2}
\tag{A.1}
\end{equation}

where $\nu_{1,2}$ and $S_{1,2}$ refer to two frequencies used to calculate the spectral index and flux densities respectively. The uncertainty in VLBI flux density measurements comprises two primary components: a random error, approximately equivalent to the image noise, and a systematic error due to visibility amplitude calibration, typically ranging from 5\% to 10\% for VLBA \citep{2009AJ....138.1874L}. This systematic error, arising from uncertainties in antenna gain curves, opacity corrections, and other factors, often dominates the overall error budget for bright sources.

\begin{figure*}
    \centering
    \includegraphics[width=0.8\textwidth]{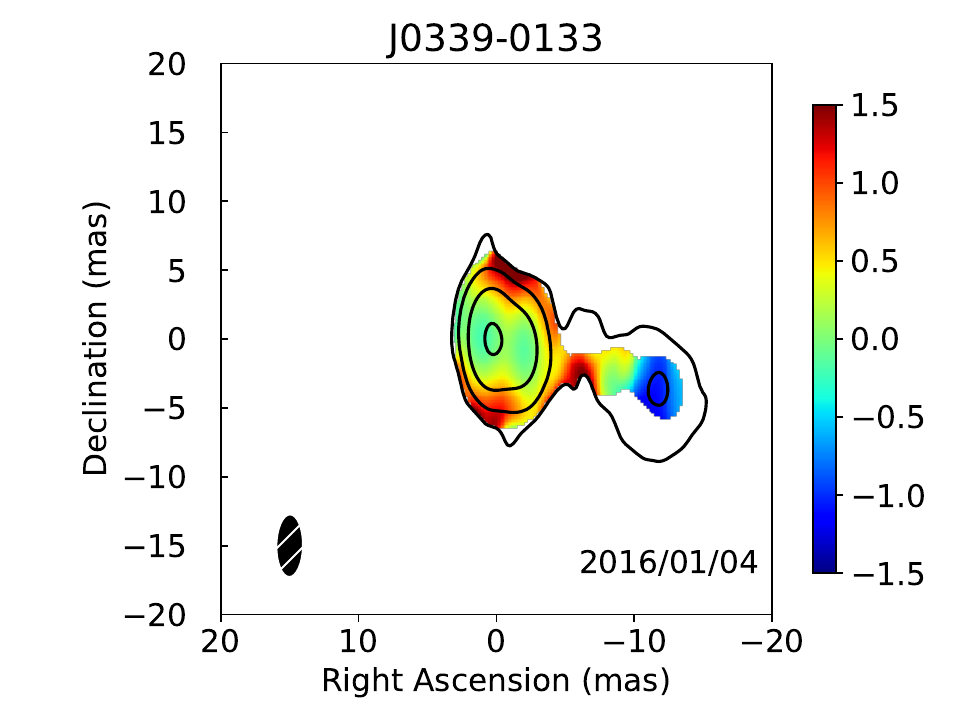}
    \includegraphics[width=0.8\textwidth]{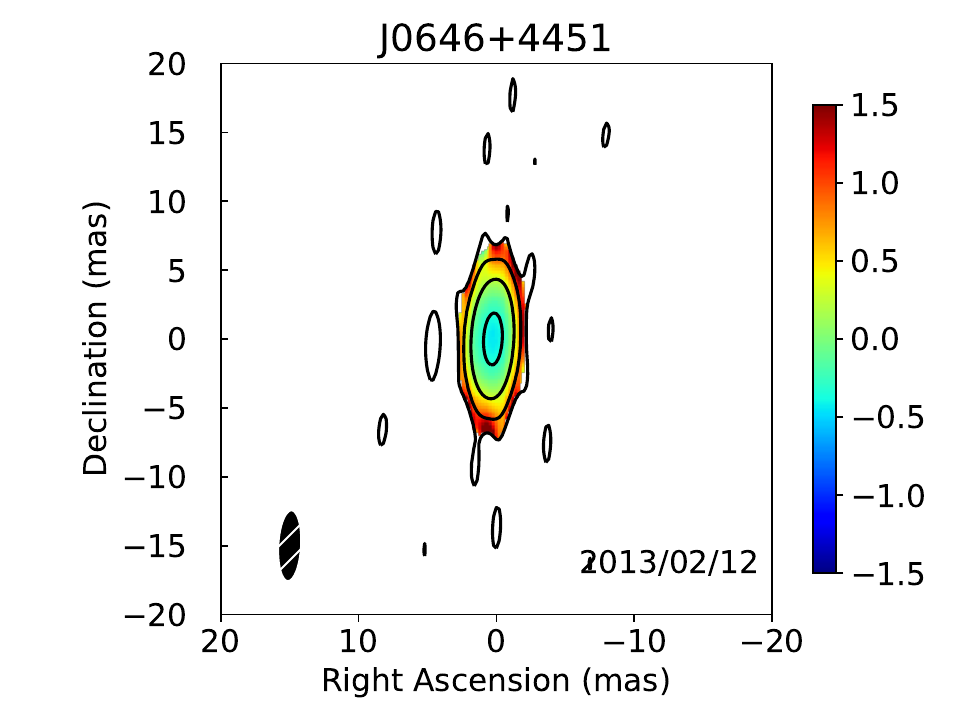}
    \includegraphics[width=0.8\textwidth]{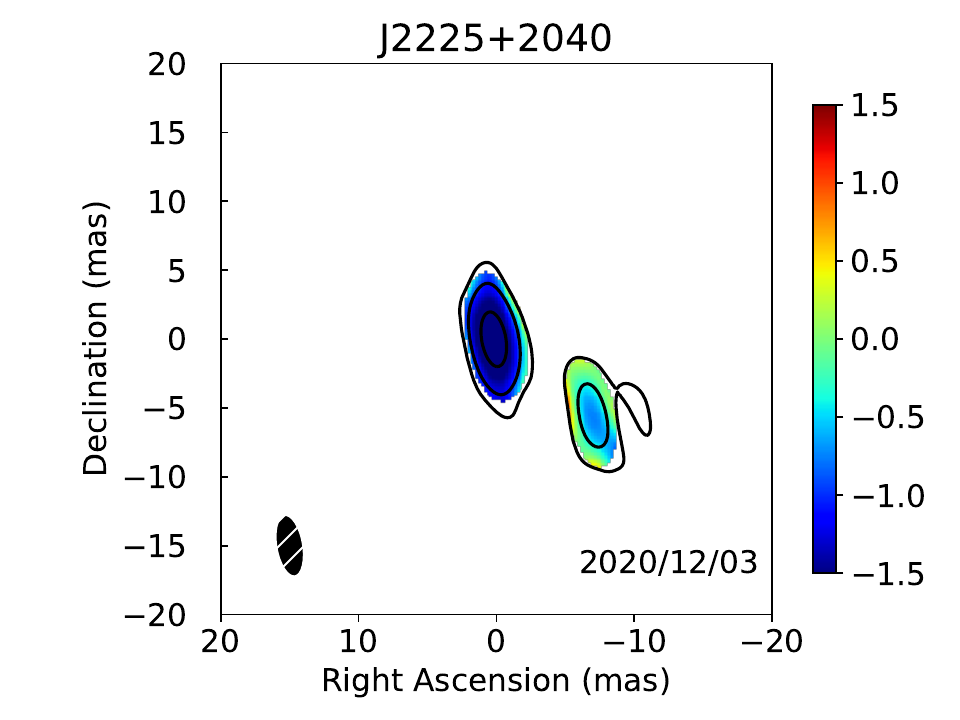}
    \caption{Example of the spectral index maps of selected sources: core-jet, core, double jet.}
    \label{fig:vlbi:indexmap}
\end{figure*}

\PublishersNote{}
\end{adjustwidth}
\end{document}